%% file: ms.tex
\documentclass[iop]{emulateapj}
\usepackage[citebordercolor={0 .5 .5}]{hyperref}
\usepackage{multirow}

\newcommand{\beq}{\begin{equation}}
\newcommand{\eeq}{\end{equation}}
\newcommand{\be}{\begin{equation}}
\newcommand{\ee}{\end{equation}}
\newcommand{\bea}{\begin{eqnarray}}
\newcommand{\eea}{\end{eqnarray}}
\newcommand{\bdi}{\begin{displaymath}}
\newcommand{\edi}{\end{displaymath}}

\newcommand{\rmicron}{$\,\micro$m}
\newcommand{\simstack}{{\sc simstack}}

\def\lsim{\,\lower2truept\hbox{${<\atop\hbox{\raise4truept\hbox{$\sim$}}}$}\,}
\def\gsim{\,\lower2truept\hbox{${>\atop\hbox{\raise4truept\hbox{$\sim$}}}$}\,}

\usepackage{amssymb,amsmath}
\usepackage{natbib}  
\usepackage[mediumspace,Gray,squaren]{SIunits}  
\shorttitle{Relating the Cosmic Optical and Infrared Backgrounds}

\shortauthors{Viero, M.\,P. et al.}
\begin{document}
\title{HerMES: The Contribution to the Cosmic Infrared Background from Galaxies Selected by Mass and Redshift$^{\dagger}$}
\author{M.P.~Viero\altaffilmark{1,$\ddagger$},
L.~Moncelsi\altaffilmark{1},
R.F.~Quadri\altaffilmark{2,3},
V.~Arumugam\altaffilmark{4},
R.J.~Assef\altaffilmark{5,6,7},
M.~B{\'e}thermin\altaffilmark{8,9},
J.~Bock\altaffilmark{1,6},
C.~Bridge\altaffilmark{1},
C.M.~Casey\altaffilmark{10},
A.~Conley\altaffilmark{11},
A.~Cooray\altaffilmark{12,1},
D.~Farrah\altaffilmark{13},
J.~Glenn\altaffilmark{14,11},
S.~Heinis\altaffilmark{15},
E.~Ibar\altaffilmark{16,17},
S.~Ikarashi\altaffilmark{18},
R.J.~Ivison\altaffilmark{17,4},
K.~Kohno\altaffilmark{18,19},
G.~Marsden\altaffilmark{20},
S.J.~Oliver\altaffilmark{21},
I.G.~Roseboom\altaffilmark{4},
B.~Schulz\altaffilmark{1,22},
D.~Scott\altaffilmark{20},
P.~Serra\altaffilmark{9},
M.~Vaccari\altaffilmark{23},
J.D.~Vieira\altaffilmark{1},
L.~Wang\altaffilmark{24},
J.~Wardlow\altaffilmark{12},
G.W.~Wilson\altaffilmark{25},
M.S.~Yun\altaffilmark{25},
M.~Zemcov\altaffilmark{1,6}}
\altaffiltext{$\ddagger$}{Email: marco.viero@caltech.edu}
\altaffiltext{$\dagger$}{Herschel is an ESA space observatory with science instruments provided by European-led Principal Investigator consortia and with important participation from NASA.} 
\altaffiltext{1}{California Institute of Technology, 1200 E. California Blvd., Pasadena, CA 91125}
\altaffiltext{2}{Carnegie Observatories, Pasadena, CA 91101}
\altaffiltext{3}{Hubble Fellow}
\altaffiltext{4}{Institute for Astronomy, University of Edinburgh, Royal Observatory, Blackford Hill, Edinburgh EH9 3HJ, UK}
\altaffiltext{5}{N\'ucleo de Astronom\'ia de la Facultad de Ingenier\'ia, Universidad Diego Portales, Av. Ej\'ercito Libertador 441, Santiago, Chile}
\altaffiltext{6}{Jet Propulsion Laboratory, 4800 Oak Grove Drive, Pasadena, CA 91109}
\altaffiltext{7}{NASA Postdoctoral Program Fellow}
\altaffiltext{8}{Laboratoire AIM-Paris-Saclay, CEA/DSM/Irfu - CNRS - Universit\'e Paris Diderot, CE-Saclay, pt courrier 131, F-91191 Gif-sur-Yvette, France}
\altaffiltext{9}{Institut d'Astrophysique Spatiale (IAS), b\^atiment 121, Universit\'e Paris-Sud 11 and CNRS (UMR 8617), 91405 Orsay, France}
\altaffiltext{10}{Institute for Astronomy, University of Hawaii, 2680 Woodlawn Drive, Honolulu, HI 96822}
\altaffiltext{11}{Center for Astrophysics and Space Astronomy 389-UCB, University of Colorado, Boulder, CO 80309}
\altaffiltext{12}{Dept. of Physics \& Astronomy, University of California, Irvine, CA 92697}
\altaffiltext{13}{Department of Physics, Virginia Tech, Blacksburg, VA 24061}
\altaffiltext{14}{Dept. of Astrophysical and Planetary Sciences, CASA 389-UCB, University of Colorado, Boulder, CO 80309}
\altaffiltext{15}{Laboratoire d'Astrophysique de Marseille - LAM, Universit\'e d'Aix-Marseille \& CNRS, UMR7326, 38 rue F. Joliot-Curie, 13388 Marseille Cedex 13, France}
\altaffiltext{16}{Departamento de Astronom\'ia y Astrof\'isica, Pontificia Universidad Catolica de Chile, Vicu\~na Mackenna 4860, Casilla 306, Santiago 22, Chile}
\altaffiltext{17}{UK Astronomy Technology Centre, Royal Observatory, Blackford Hill, Edinburgh EH9 3HJ, UK}
\altaffiltext{18}{ Institute of Astronomy, University of Tokyo, 2-21-1 Osawa, Mitaka, Tokyo 181-0015, Japan}
\altaffiltext{19}{ Research Center for the Early Universe, University of Tokyo, 7-3-1 Hongo, Bunkyo, Tokyo 113-0033, Japan}
\altaffiltext{20}{Department of Physics \& Astronomy, University of British Columbia, 6224 Agricultural Road, Vancouver, BC V6T~1Z1, Canada}
\altaffiltext{21}{Astronomy Centre, Dept. of Physics \& Astronomy, University of Sussex, Brighton BN1 9QH, UK}
\altaffiltext{22}{Infrared Processing and Analysis Center, MS 100-22, California Institute of Technology, JPL, Pasadena, CA 91125}
\altaffiltext{23}{Astrophysics Group, Physics Department, University of the Western Cape, Private Bag X17, 7535, Bellville, Cape Town, South Africa}
\altaffiltext{24}{Institute for Computational Cosmology, Department of Physics, University of Durham, South Road, Durham, DH1 3LE, UK}
\altaffiltext{25}{Department of Astronomy, University of Massachusetts, Amherst, MA 01003}
\begin{abstract}
We quantify the fraction of the cosmic infrared background (CIB) that originates from galaxies identified in the  UV/optical/near-infrared by stacking 81{,}250 ($\sim 35.7\, \rm arcmin^{-2}$) \emph{K}-selected sources ($K_{\rm AB}<24.0$)  split according to their rest-frame $U-V$ vs.\@ $V-J$ colors into 72{,}216 star-forming and 9{,}034  quiescent galaxies, on maps from \emph{Spitzer}/MIPS (24\rmicron), \emph{Herschel}/PACS (100, 160\rmicron),  \emph{Herschel}/SPIRE (250, 350, 500\rmicron), and AzTEC (1100\rmicron).  
The fraction of the CIB resolved by our catalog is ($69 \pm 15$)\% at 24\rmicron,   ($78 \pm 17$)\% at 70\rmicron,  ($58 \pm 13$)\% at 100\rmicron, ($78 \pm 18$)\% at 160\rmicron,   ($80 \pm 17$)\% at 250\rmicron, ($69 \pm 14$)\% at 350\rmicron,  ($65 \pm 12$)\% at 500\rmicron, and ($45 \pm 8$)\% at 1100\rmicron.  Of that total, about 95\% originates from star-forming galaxies, while the remaining 5\% is from apparently quiescent galaxies.  
The CIB at $\lambda \lsim 200$\rmicron\ appears to be sourced predominantly from galaxies at $z \lsim 1$, while at $\lambda \gsim 200$\rmicron\ the bulk originates from  
$ 1 \lsim z \lsim 2$.  Galaxies with stellar masses log($M/ \rm M_{\odot})=9.5$--11 are responsible for the majority of the CIB,  with those in the log($M/ \rm M_{\odot})=9.5$--10 bin contributing mostly at $\lambda < 250$\rmicron, and those in the log($M/ \rm M_{\odot})=10$--11 bin dominating at $\lambda > 350$\rmicron.   
The contribution from galaxies in the  log($M/ \rm M_{\odot})=9.0$--9.5 (lowest) and log($M/ \rm M_{\odot})=11.0$--12.0 (highest) stellar-mass bins contribute the least---both of order 5\%---although the highest stellar-mass bin is a significant contributor to the luminosity density at $z\gsim 2$.      
The luminosities of the galaxies responsible for the CIB shifts from combinations of \lq\lq normal\rq\rq\ and luminous infrared galaxies (LIRGs) at $\lambda \lsim 160$\rmicron, to LIRGs at $160 \lsim \lambda \lsim 500$\rmicron, to finally LIRGs and ultra-luminous infrared galaxies (ULIRGs) at $\lambda \gsim 500$\rmicron.  
Stacking analyses were performed using \simstack, a novel algorithm designed to account for possible biases in the stacked flux density due to clustering.  It is made available to the public at \url{www.astro.caltech.edu/~viero/viero_homepage/toolbox.html}.  
\end{abstract}


% ======================
\section{Introduction}
\label{sec:intro}
% ======================
 The cosmic infrared background (CIB), discovered in Far Infrared Absolute Spectrophotometer (FIRAS) data from the \emph{Cosmic Background Explorer}  \citep[\emph{COBE;}][]{puget1996,fixsen1998}, originates from 
thermal re-radiation of UV/optical starlight (and to a lesser extent active galactic nuclei, or AGN, emission) absorbed by dust grains.  
The total intensity of this background is roughly equal to that of the combined extragalactic 
UV, optical, and near-infrared backgrounds (the \lq\lq cosmic optical background\rq\rq, or COB) 
indicating that, of all the light ever emitted by stars, about half has been absorbed and re-emitted by dust \citep{hauser2001}. 
While it is thought that the majority of the CIB originates from dusty star-forming galaxies \citep[DSFGs; e.g.,][]{lefloch2005,lagache2005,viero2009},  how they relate to the sources that make up the COB, and 
what fraction of the CIB is resolvable as optical sources, is still unclear.  

\begin{figure}[!t]
\centering
\includegraphics[width=0.45\textwidth]{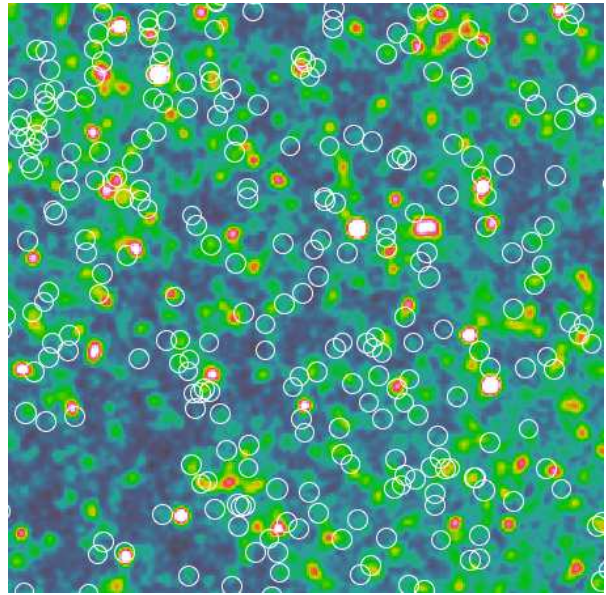}
\caption{$0.25^{\circ} \times 0.25^{\circ}$ SPIRE 250\rmicron\ cutout of the {\sc UDS} field, which has been smoothed and color-stretched to visually enhance the regions of submillimeter emission.  Overlaid as white circles (diameters of 30\arcsec) are the positions of star-forming galaxies with masses between $\sim 10^{9.5-10.0}\, \rm M_{\odot}$, in a single redshift slice spanning $z= 1.0$--1.5.  Note that very few, if any, of the \emph{K}-selected sources are resolved in the SPIRE map, but that most lie on ridges of faint emission.}
\label{fig:thumb}
\end{figure}

To definitively answer that question, ideally the CIB would be resolved into individual sources and matched to optical counterparts,   
but from the first DSFGs imaged  in the submm \citep[e.g.,][]{smail1997, barger1998, hughes1998} it became quickly evident that identifying optical counterparts is a non-trivial exercise. 
The angular resolution afforded by single-dish submillimeter observatories results in beams containing multiple sources, 
such that in deep observations the spatial variation of the sky intensity eventually reaches the so-called \lq\lq confusion limit\rq\rq\ \citep[e.g.,][]{nguyen2010}.   
This situation is made worse by the strong evolution undergone by DSFGs between the present day and $z\sim 1$ \citep[e.g.,][]{pascale2009},  such that only the brightest $\sim 1\%$ of DSFGs \citep[equivalent to $\sim 15\% $ of the CIB; ][]{oliver2010} at 250\rmicron\ is resolvable into point sources.  
This is illustrated in Figure~\ref{fig:thumb}, where a $0.25 \times 0.25\, \rm deg^2$ cutout of the SPIRE 250\rmicron\ map is overlaid with positions of star-forming galaxies with masses between $\sim 10^{9.5-10.0}\, \rm M_{\odot}$, at $z= 1.0$--1.5.  
The map is smoothed and color-stretched to highlight the regions of emission.  It is clear that very few of the sources are detected individually---the rest lie almost exclusively on ridges of faint emission.  

%%%%%%%%%%%%%%%%%%%%
%%%%%%%%%%%%%%%%%%%%

Given ancillary data of sufficient quality, this limitation can be overcome by stacking.  
Conceptually, stacking is very simple: imagine cutting out hundreds of thumbnails from a  map centered on the positions where galaxies are known to be, and averaging those thumbnails together until an image of the average galaxy emerges from the noise.  
These positional priors can come in many forms, e.g., they could be catalogs of UV, optical, IR or radio sources.   
Note that the output is the \emph{average} of that population in the stacked maps, i.e., there will likely be sources whose actual fluxes are higher or lower.  
Thus, the more homogeneous the sources comprising the input list, the more meaningful the stacked flux will be.  
 
Stacking has been successfully applied to infrared maps by numerous groups looking to resolve the CIB with resolved sources.  
Frequently, Multi-band Imaging Photometer for \emph{Spitzer} \citep[MIPS;][]{rieke2004} 24\rmicron\ sources were used as positional priors because bright 24\rmicron\ sources are usually correlated with far-infrared (FIR) and submillimeter emission, and because of the large fraction of sources that are resolved in 24\rmicron\ maps \citep[$\sim 70\%$;][]{papovich2004}.   
For example, \citet{dole2006} showed that much of the CIB at 70 and 160\rmicron\ is resolved by sources  whose flux densities at 24\rmicron\  are $S \gsim 60\, \micro \rm Jy$.   Similarly, by stacking on maps from the Balloon-borne Large-Aperture Submillimeter Telescope \citep[BLAST; ][]{pascale2008},  \citet{devlin2009} and \citet{marsden2009} demonstrated that close to the full intensity of the CIB at 250, 350, and 500\rmicron\ is resolved by 24\rmicron\ sources, and that roughly half of the CIB at 500\rmicron\ originates at $z > 1.2$.  
\citet{jauzac2011} estimated the contribution to the CIB brightness at 70 and 160\rmicron\ from 24\rmicron\ sources by stacking in narrow redshift bins spanning $0<z<1.05$.  
\citet{berta2011}, using PACS, and \citet{bethermin2012b}, using SPIRE maps, reconstructed number counts to flux densities below the confusion limit by stacking with 24\rmicron\ priors, and found that the integral of their counts resolved 58--74\%,  
and 55--73\% of the CIB at their respective wavelengths.  
And \citet{penner2011} stacked  24\rmicron\ sources in AzTEC (1.1\,mm) and MAMBO (1.2\,mm) data, finding much of the resolved background at those wavelengths originates from $z>1.3$.  

Stacking has been used to address other questions as well.  
\citet{pascale2009}, stacking 24\rmicron\ sources on BLAST maps, measured the evolution of the infrared luminosity density with redshift, finding  a significant rise in the temperatures and luminosities of sources with increasing redshift.  
\citet{oliver2010b} used \emph{Spitzer} \lq\lq bandmerged\rq\rq\ catalogs to measure the mass-dependency of specific star formation rates (sSFR)---the star-formation rate of the galaxy divided by its mass---over the redshift range $0  < z < 2$.  
 \citet{viero2012} stacked high-redshift massive galaxies from the GOODS-NICMOS Survey \citep[GNS;][]{conselice2011} on maps from PACS at 70--160\rmicron\ \citep{poglitsch2010}, 870\rmicron\ from LABOCA \citep{weiss2009}, and BLAST, finding that the bulk of the star formation occurs in disk-like galaxies, with a hint that spheroid-like galaxies harbor a low level of star formation as well.   
Similarly, \citet{hilton2012} stacked a stellar-mass selected sample of $1.5 < z < 3$ galaxies drawn from the GNS on SPIRE maps and found evidence for an increasing fraction of dust-obscured star formation with stellar mass.  
And \citet{heinis2012} stacked ultraviolet selected galaxies at $z\sim 1.5$ on SPIRE maps, finding that the mean infrared luminosity is correlated to the slope of the UV continuum, $\beta$.

While conceptually simple,  in practice proper stacking is subtle and not without controversy.  
For sources that are uncorrelated (i.e., not clustered around other galaxies) the technique returns an unbiased estimate of the average flux density \citep[e.g.,][]{marsden2009,viero2012}.  
 But if sources \emph{are} clustered---which they inevitably will be---then a bias at some level will be present and must be accounted for. 

Many solutions have been proposed to address this problem:  some, like \citet{bethermin2012b}  correct for boosting with simulations.    
Alternatively,  \citet{bethermin2012b} and \citet{heinis2012} fit the measured correlation function to the excess width of the measured stacked beam to estimate a correction.  
And \citet{bourne2011} use a median statistic to perform their stacking, which is shown to be resistant to biases induced by outliers.   

Still another method, developed independently by \citet{kurczynski2010}, \citet{roseboom2010}, and \citet{bourne2012}, 
fits for the flux densities of multiple (correlated) lists simultaneously, thereby accounting for correlations as a part of the stack.  
The advantage of this technique is that it makes few assumptions  and naturally takes into account the possibility that the potential clustering bias may be redshift and luminosity dependent.   Here we build upon this technique to simultaneously measure the mean flux densities of galaxies selected by mass and divided into mass and redshift bins.

Our goal is to gain a better understanding of the contribution to the CIB from galaxies identified in the optical and near-infrared.  
In \S~\ref{sec:data} we present the data, and in \S~\ref{sec:formalism} we present our method,  demonstrating its effectiveness with simulations in \S~\ref{sec:simulations}.  We ultimately use it  to determine the total contribution to the CIB from \emph{K}-selected galaxies (\S~\ref{sec:resolved_cib}), and its dependence on redshift (\S~\ref{sec:z}), mass and color (\S~\ref{sec:mass}), and luminosity (\S~\ref{sec:luminosity}).  
The dependence of sSFR on these variables will be explored in a forthcoming paper (Arumugam et al.\@ in prep.).  

When required, we assume a \citet{chabrier2003} initial mass function (IMF) and a flat $\Lambda$CDM cosmology with $\Omega_{\rm M} = 0.274$, $\Omega_{\Lambda} = 0.726$, $H_0 = 70.5\, \rm km\, s^{-1}\, Mpc^{-1}$, and $\sigma_8 = 0.81$ \citep{komatsu2011}.  

% ======================
\section{Data}
\label{sec:data}
% ======================
We perform our analysis on the UKIRT Infrared Deep Sky Survey \citep[UKIDSS;][]{lawrence2007},  Ultra-Deep Survey 
 (UDS)  field, centered at coordinates $\rm 2^{h}17^m50^s, -5^{\circ}6\arcmin 0\arcsec$.
The {\sc UDS} is the deepest survey undertaken by UKIDSS, covering $0.8\, \rm deg^2$ in $J$, $H$, and $K$ to nominal 5$\sigma$ depths of 26.9, 25.9, 24.9\,mag\,[AB].
Catalogs are based on optical and near-infrared (NIR) data in this field, while the maps on which the stacking analyses are performed span the mid-infrared to submillimeter.
Here we briefly describe the catalog and maps.
%#######################
\subsection{Optical/Near-Infrared Catalog}
\label{sec:catalog}
% #######################
%%%%%%%%%%%%%%%%%%%%
\begin{figure}%[!t]
\centering
\vspace{-4.0mm}
\hspace{-10mm}
\includegraphics[width=0.50\textwidth]{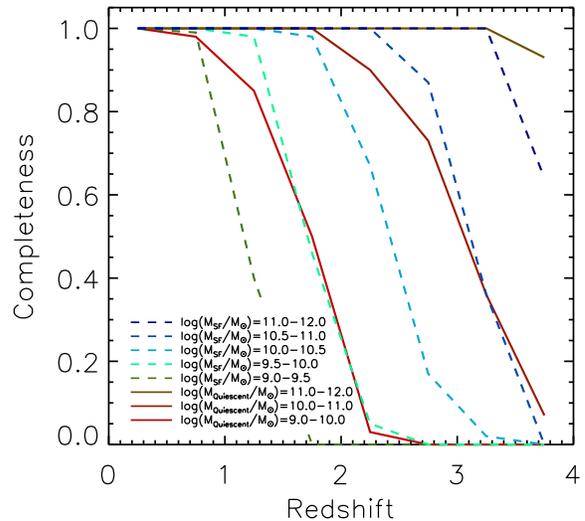}
\caption{Completeness estimates plotted vs.\@ redshift for star-forming and quiescent galaxies in bins of stellar mass.}
\label{fig:comp}
\end{figure}
%%%%%%%%%%%%%%%%%%%%
Galaxy positions and redshifts come from a catalog based on the UKIDSS
Ultra-Deep Survey \citep[{\sc UDS}; ][]{lawrence2007, warren2007} Data Release 8 and
supplementary data (Williams et al. 2013, in prep.).  Source detection,
photometry, and spectral energy distribution (SED) fitting are described in detail by \citet{williams2009},
with recent updates to the catalog discussed by \citet{quadri2012}.  A brief
summary of the data follows:  
sources are detected in the {\sc UDS} \emph{K}-band mosaic
using \emph{Source Extractor} v2.5.0 \citep*{bertin1996}, and fluxes measured
in other bands from PSF-matched images: $u^{\prime}$ from archival CFHT
data, $BVRi^{\prime} z^{\prime}$ from the Subaru-XMM Deep Survey
\citep[SXDS;][]{furusawa2008}, $JHK$ from the {\sc UDS} DR8, 
and \emph{Spizter}/IRAC imaging at 3.6, 4.5, 5.8, and 8\rmicron\ in the {\sc UDS} field.  
An inspection of the number counts suggests that our catalog is essentially complete to $K_{\rm AB}<24.0$ for non-stellar objects, 
although it is difficult to rule out the possibility that there are a small number of diffuse and extended sources that do not make it into our sample.   
The de-blending technique of \citet{labbe2006} and \citet{wuyts2007} is used to extract matched IRAC fluxes.  
Objects near bad pixels in the optical or near-infrared images, or those with no optical coverage, are excluded, as well as stars, saturated sources, severe blends, or those near the edge of the image.  These quality checks reduce the catalog from 171{,}392 to 81{,}250 objects, and the resulting effective image area is $\sim 0.63\, \rm deg^2$.  

Photometric redshifts and rest-frame colors are derived by fitting the
multi-band photometry with {\sc EAZY} \citep{brammer2008}.  Stellar
masses are obtained with {\sc FAST} \citep{kriek2009a} using a
\citet{chabrier2003} IMF, solar metallicity,
and \citet{bruzual2003} stellar population models.

Quiescent and star-forming galaxies are classified based on the observed bimodality in a rest-frame color-color $U-V$
vs.\@ $V-J$ \citep[hereafter \emph{UVJ};][]{williams2009}; this technique robustly separates red, dusty starbursts from red,
dust-free, old stellar populations \citep[see also ][]{labbe2005,wuyts2007}. In addition to the quality checks listed above, sources whose best-fit SED has a high $\chi^2$ ---which may be the result of poor photometry, artifacts in the images, or of a strong active galactic nucleus (AGN) component that is not a part of the {\sc EAZY} template library--- are excluded from the parent sample.  
Approximately 2{,}700 sources ($\sim 3\%$) exceed this limit;  we explore the effect that inclusion of this subsample has on the total resolved CIB in \S~\ref{sec:agn}.   

Our sample is selected at $K_{\rm AB} \le 24$, where the catalog is essentially 100\% complete.  We calculate the corresponding mass completeness values in a manner similar to  \citet{quadri2012}.  Briefly, we scale the fluxes and masses of galaxies at slightly brighter magnitudes down to $K_{\rm AB} = 24$, and estimate the completeness as a function of mass and redshift as the fraction of objects below that mass at that redshift.  Completeness estimates  are plotted in Figure~\ref{fig:comp}.

%%%%%%%%TABLE 1%%%%%%%%%
\input{table1.tex}
%%%%%%%%TABLE 1%%%%%%%%%

%#######################
\subsection{Spitzer/MIPS}
\label{sec:mips}
% #######################
We use publicly available   \emph{Spitzer}/MIPS maps at 70, and 160\rmicron\ from the  Spitzer Wide-Area Infrared Extragalactic (SWIRE) survey \citep{lonsdale2003} in the XMM Large-Scale Structure field   \citep[{\sc xmm-lss;}\rm][]{surace2005}; 
and at 24\rmicron\ from the \emph{Spitzer} UDS survey (SpUDS; PI: J.~Dunlop), DR2.
Maps have RMS levels of 0.5\,mJy, 1.8\,mJy and 19.9\,$\micro$Jy, respectively.    
We note that the absolute calibration uncertainties are 4, 7, and 12\% at 24, 70, and 160\rmicron, respectively \citep{engelbracht2007, gordon2007, stansberry2007}.

Following  \citet{bethermin2010}, calibration corrections of 1.0509, 1.10, and 0.98, and aperture corrections of 1.19, 1.21, and 1.20, are applied at 24, 70, and 160\rmicron.  
Maps are in native units of $\rm MJy\, sr^{-1}$ (surface brightness), and are converted to $\rm Jy\, beam^{-1}$ for this analysis by dividing the maps by the solid angles of the measured instrumental point response functions\footnote[1]{\url{http://irsa.ipac.caltech.edu/data/SPITZER/docs/mips/}}  (PRFs), where $\Omega_{\rm beam} = f/I_0 = \int \rm PRF\, d\Omega/PRF_0$, and $\rm PRF_0$ is the peak value.   

Also measured from the PRF is the effective full width at half maximum (FWHM) of the best-fit Gaussian, which can differ from nominal by as much as $\sim 6\%$.  This is done by simply finding the 2D Gaussian which provides the minimum value when differenced with the  PRF, within a radius of $1.25\times \rm FWHM$ (chosen as the approximate minimum of the primary lobe).  Effective area and FWHM for each band are listed in Table~\ref{tab:beams}.  

%#######################
\subsection{Herschel/HerMES}

We use submillimeter maps at 100, 160, 250, 350, and 500\rmicron\ from the the Herschel Multi-tiered Extragalactic Survey \citep[HerMES;][]{oliver2012}.
HerMES is a guaranteed time (GT) key project, and consists of maps of many of the well studied extragalactic fields, which are divided into tiers of depth and area, observed with both the Spectral and Photometric Imaging REceiver \citep[SPIRE][]{griffin2010} and the Photodetector Array Camera and Spectrometer \citep[PACS][]{poglitsch2010}.  The {\sc UDS} is a level 4 field, consisting of 20 repeat observations, and will be made available via {\sc HeDaM}\footnote[2]{\url{http://hedam.oamp.fr/HerMES/}} \citep{roehlly2011}, as part of DR2. 
  
\subsubsection{Herschel/PACS}
\label{sec:pacs}
% #######################
Data at 100 and 160\rmicron\ are taken with the PACS instrument and  are processed with the 
Herschel Data Processing System \citep[HIPE v10.2747;][]{ott2010}.   
Maps were made with {\sc UniHIPE}\footnote[3]{\url{herschel.asdc.asi.it/index.php?page=unimap.html}} 
in combination with {\sc Unimap}\footnote[4]{\url{w3.uniroma1.it/unimap}} \citep{traficante2011,piazzo2013}.   
Maps are made using the default parameters, with the exception of the image pixel sizes, which we set to 2 and 3 arcsec at 100 and 160\rmicron , respectively.   
The advantage of this mapmaker over the standard one available through HIPE is that it does not require strong high-pass filtering or masking of bright sources \citep[e.g.,][]{wieprecht2009} to produce reliable maps, thus avoiding the attenuation of the fainter population as was found in \citep[e.g.,][]{lutz2011,viero2012}.
 The r.m.s.\@ depths of the maps are 0.44 and 1.5\,mJy at 100 and 160\rmicron, respectively.  

%#######################
\subsubsection{Herschel/SPIRE}
\label{sec:spire}
% #######################
Data at 250, 350, and 500\rmicron\ are observed with the SPIRE instrument  to a depth of 11.2, 9.3, 13.4\,mJy (5$\sigma$), not including confusion noise, which from \citet{nguyen2010} is 24.0, 27.5, and 30.5\,mJy (5$\sigma$) at 250, 350 and 500\rmicron, respectively.  
Absolute calibration is detailed in \citet{swinyard2010}, with calibration uncertainties of $\sim 7\%$.   
Maps are made  with 3 arcsec pixels using {\sc SMAP} \citep{levenson2010, viero2013a}.  
%#######################
\subsection{AzTEC}
\label{sec:aztec}
% #######################
We use maps at 1100\rmicron\ observed 
with the AzTEC camera \citep{wilson2008, glenn1998}  mounted on ASTE \citep{ezawa2004,ezawa2008}.  The FWHM of the AzTEC beam on ASTE is 30\,arcsec at 1100\rmicron, and the field of view of the array is roughly circular with a diameter of 8 arcmin.   
Calibration errors are quoted for individual observations to be 6--13\% \citep{wilson2008,austermann2010}, depending on the source; here we adopt a value of 10\%.  
The area covered is smaller than at the other bands, totaling $\sim 0.32\, \rm deg^2$ after cropping the noisy outer edge. 
These data will be presented by Ikarashi et al.\@ (in prep.).  

%#######################
\subsection{Color Corrections}
\label{sec:color_correction}
 %#######################
We apply color corrections to convert from the standard calibration  to the actual measured SED of the stacked sources.  
As the part of the spectrum observed depends on the source's redshift, the color correction is applied after first finding the best-fit SED in each bin.  
Consequently, each color correction is unique, though the difference in any one band across the full redshift range is never greater than $\sim 10\%$.  
The color corrections per band, from lowest to highest redshift, are: 0.99--1.02 (24\rmicron);  0.97--1.02 (70\rmicron);  0.93--1.02 (100\rmicron);  0.99--1.00 (160\rmicron);  0.98--0.99 (250\rmicron);  0.99--1.00 (350\rmicron);  0.99--1.07 (500\rmicron); and  0.96--0.99 (1100\rmicron).  
 % ======================
\section{Method for Unbiased Stacking}
\label{sec:method}
% ======================
As was shown by \citet{marsden2009}, stacking is formally the covariance between a catalog (or multiple catalogs) of positions $C_{\alpha}$, containing $N_{\alpha}^j$ sources in pixel $j$, and a map, $M$.  The mean of $N_{\alpha}^j$ is $\mu_{\alpha}$, which represents the average number of sources in catalog $\alpha$ per pixel.   
In the limit that sources are Poisson distributed \emph{on the scale of the beam}, then the covariance is simply the mean of the map at positions $C_{\alpha}$,  so that the average flux density of a given catalog is
\beq
\hat{S}_{\alpha}= \frac{1}{N_{\rm pix}\mu_{\alpha}}\sum_j M_j N^j_{\alpha},
\eeq
where $N_{\rm pix}$ is the total number of pixels in the map.  

If the catalog (or catalogs) in question is correlated on the scale of the beam, $\mu$ can simply be replaced with the variance, $\sigma^2$.
What this does not account for---as pointed out by e.g., \citet[][]{chary2010}, \citet{serjeant2010}, and \citet{kurczynski2010}---is the possibility that some other, fainter, and potentially numerous sources (or the sources in companion catalogs),  may be correlated with the sources in that catalog, and that neglecting them could introduce a bias.

We now present an algorithm, whose formalism is similar to those of \citet[][]{kurczynski2010} and \citet[][]{roseboom2010}, with the difference that only samples which could potentially be correlated (i.e., those in the same redshift range) are simultaneously fit.   
In the following section we provide the formalism, while step-by-step instructions are given in \S~\ref{sec:cookbook}.  

%#######################
\subsection{Stacking Formalism}
\label{sec:formalism}
% #######################

The following is a generalization of the formalism presented in  \citet{marsden2009}, and is applicable to {\it any}  catalog, including those that are clustered at angular scales comparable to that of the beam.
For a map, $M_j$, with pixels $j$, and a set of lists, $S_{\alpha}$:  
\begin{align}
\label{eqn:simultaneous_1}
M_j&= n_j + \sum_{\alpha} S_{\alpha} \left(N_{\alpha}^j
-\mu_{\alpha}\right) \nonumber \\
&= n_j + S_1 \left(N_1^j -\mu_1\right)+  
\ldots + S_n \left(N_n^j -\mu_n\right),
\end{align}
where the $S_{\alpha}$ form the complete set of all
objects in the Universe.

Note that, unlike in \citet{marsden2009}, we need not assume that $N_{\alpha}^j$ be a Poisson-distributed number.  
Furthermore, separate lists can also be correlated, so that the covariances between them need not be non-zero.  
However, we still require that the instrumental noise is well behaved, i.e., $\left<
n_j \right>=0$, so that terms in $N_{\alpha}^j\,n_j$ vanish in the
sum.

The amplitudes 
$S_{\alpha}$ and $N_{\alpha}^j$  
in Equation~\ref{eqn:simultaneous_1} 
that satisfy $M_j$ can be quantified by
writing their covariances with the map itself:
\begin{align}
\label{eqn:simultaneous_2}
{\rm Cov}(M,N_{\alpha})  &= \frac{1}{N_{\rm pix}}{\sum_j M_j
N_{\alpha}^j}  \nonumber \\ 
& =   \frac{1}{N_{\rm pix}}{\sum_j
N_{\alpha}^j}\sum_{\alpha} S_{\alpha} \left(N_{\alpha}^j
-\mu_{\alpha}\right)\nonumber \\
&= \frac{S_{\alpha}}{N_{\rm pix}} \left[ \sum_j
\left(N_{\alpha}^j\right)^2 - \mu_{\alpha} \sum_j N_{\alpha}^j
\right] \nonumber \\ & +  \sum_{\alpha\prime\neq\alpha}\frac{S_{\alpha\prime}}{N_{\rm
pix}} \left[ \sum_j N_{\alpha}^j N_{\alpha\prime}^j -
\mu_{\alpha\prime} \sum_j N_{\alpha}^j \right],
\end{align}
%
%Equation~\ref{eqn:simultaneous_2} 
which can be re-written in matrix form
by defining amplitude and covariance vectors:
\begin{equation}
\label{eq:simultaneous_3}
%\small
{\bf S} =
\begin{pmatrix}
 S_1\\
 S_2\\
 \vdots\\
 S_n\\
\end{pmatrix};~
{\rm \bf Cov}(M,N_{\alpha}) =
\begin{pmatrix}
 {\rm Cov}(M,N_1)\\
 {\rm Cov}(M,N_2)\\
 \vdots\\
 {\rm Cov}(M,N_n)\\
\end{pmatrix};
\end{equation}
and 
%
%\begin{equation}
\begin{align}
\label{eq:simultaneous_4}
\small
& 
\begin{pmatrix}
{\rm Cov}(M,N_1)\\
 %{\rm Cov}(M_j,N_2^j)\\
 \vdots\\
 {\rm Cov}(M,N_n)\\
\end{pmatrix}
= \frac{1}{N_{\rm pix}}
 \times  \nonumber \\ 
& 
\begin{pmatrix}
  \sum\limits_{j} N_1^j (N_1^j -\mu_1)  & \cdots & \sum\limits_{j} N_1^j (N_n^j -\mu_n) \\
  \vdots  & \ddots & \vdots \\
  \sum\limits_{j} N_n^j (N_1^j -\mu_1)   & \cdots & \sum\limits_{j} N_n^j (N_n^j -\mu_n) \\
\end{pmatrix}
\cdot
\begin{pmatrix}
 S_1\\
 \vdots\\
 S_n\\
\end{pmatrix}
.
\end{align}

From the covariances and the $n\times n$ matrix which we label 
{\bf A},  {\bf S} is then simply
\begin{equation}\label{eq:simultaneous_5}
    {\rm \bf \tilde{S}} = {\bf A}^{-1}~{\rm \bf Cov}(M,N_{\alpha}).
\end{equation}

Notice the resemblance that the linear system in
Equation~\ref{eq:simultaneous_4} bears  to
that of a least-squares fit 
\begin{equation}\label{eq:simultaneous_6}
y = \sum_{\alpha} a_{\alpha} x_{\alpha} = a_1 x_1 + a_2 x_2 +
\ldots + a_n x_n,
\end{equation}
whose residual is given by 
\begin{equation}\label{eq:simultaneous_7}
R^2 = \sum\limits_{j}\left[y_j - \left(a_1 x^j_1 + a_2 x^j_2 +
\ldots + a_n x^j_n\right)\right]^2.
\end{equation}

In order to minimize this residual, we impose the following set of
conditions:
\begin{eqnarray}\label{eqn:simultaneous_8}
\frac{\partial R^2}{\partial a_1} &=& -2 \sum\limits_{j}\left[y_j - \left(a_1 x^j_1 + a_2 x^j_2 + \ldots + a_n x^j_n\right)\right]x_1^j =0 \nonumber;\\
\frac{\partial R^2}{\partial a_2} &=& -2 \sum\limits_{j}\left[y_j - \left(a_1 x^j_1 + a_2 x^j_2 + \ldots + a_n x^j_n\right)\right]x_2^j =0\nonumber;\\
\vdots&~~&~~~~~~~~~~\vdots~~~~~~~~~~~~\vdots~~~~~~~~~~~~\vdots \nonumber \\
\frac{\partial R^2}{\partial a_n} &=& -2 \sum\limits_{j}\left[y_j - \left(a_1 x^j_1 + a_2 x^j_2 + \ldots + a_n x^j_n\right)\right]x_n^j =0\nonumber.\\
\end{eqnarray}
These can be expressed in matrix form as
\begin{equation}
\small
\label{eq:simultaneous_9}
\begin{pmatrix}
\sum\limits_{j} x_1^j y_j\\
\sum\limits_{j} x_2^j y_j\\
 \vdots\\
\sum\limits_{j} x_n^j y_j\\
\end{pmatrix}
=
\begin{pmatrix}
  \sum\limits_{j} (x_1^j)^2 & \sum\limits_{j} x_1^j x_2^j & \cdots & \sum\limits_{j} x_1^j x_n^j \\
  \sum\limits_{j} x_1^j x_2^j & \sum\limits_{j} (x_2^j)^2 & \cdots & \sum\limits_{j} x_2^j x_n^j \\
  \vdots & \vdots & \ddots & \vdots \\
  \sum\limits_{j} x_1^j x_n^j & \sum\limits_{j} x_2^j x_n^j & \cdots & \sum\limits_{j} (x_n^j)^2 \\
\end{pmatrix}
\cdot
\begin{pmatrix}
 a_1\\
 a_2\\
 \vdots\\
 a_n\\
\end{pmatrix}.
\end{equation}

By comparing Equations~\ref{eq:simultaneous_4} and
\ref{eq:simultaneous_9}, it is clear that the $a_{\alpha}$ vector maps 
into $S_{\alpha}$, the $y_j$ vector maps into $M_j$, the $\sum\limits_{j}
x_{\alpha}^j y_j$ vector maps into the covariances ${\rm
Cov}(M_j,N_{\alpha}^j)$, and the $x_{\alpha}^j$ vector maps into
the $N_{\alpha}^j$ (which is mean-subtracted).

Therefore, solving  Equation~\ref{eq:simultaneous_4} for $S_{\alpha}$  is
equivalent to finding the coefficients $a_{\alpha}$ in
Equation~\ref{eq:simultaneous_6} via a minimization routine.
Specifically,  the functional form can be
operatively implemented using known quantities:
\begin{equation}\label{eqn:simultaneous_10}
M = \sum_{\alpha} S_{\alpha} C_{\alpha} = S_1 C_1 + S_2 C_2+
\ldots + S_n C_n,
\end{equation}
where we define the $C_{\alpha}^j$ as a 
beam-convolved and mean-subtracted version of the $N_{\alpha}^j$.
%

%#######################
\subsection{Method in Practice}
\label{sec:cookbook}
% #######################
Here we present the simultaneous stacking algorithm (\simstack) used in this analysis, which we also make publicly available through an {\sc IDL} code\footnote[5]{\url{www.astro.caltech.edu/~viero/viero_homepage/toolbox.html}}.
The simultaneous stack is performed on one map at a time, and one group at a time, where groups are defined as catalogs which could potentially be correlated.   For example, we group all lists in the same redshift range together (for a total of 8 groups), as we expect galaxies of different masses but equal redshifts to be correlated with each other.  In other words, we assume that galaxies in different redshift slices are uncorrelated, and can be dealt with independently.  
Then, regardless of the code used, the method can be broken into four simple steps:

\begin{description}
\item[Prepare] $N$ lists of RA and Dec by group,  e.g., we divide each group (redshift slice) into 8 lists of mass and \emph{UVJ} color; thus $N=8$.
\item[Construct] $N$ layers, or \lq\lq hits\rq\rq\ maps, one for each list, where each pixel in the hits map contains the integer number of sources which falls into it. 
\item[Convolve] the $N$ layers with an effective point spread function (PSF; we use a Gaussian) whose FWHM is equal to that of the the \emph{effective} instrumental beam %(rather than the nominal one) 
of the real map\footnote[6]{If using the actual PRF of the instrument, take care that the orientation is correct, and if the field has been viewed at multiple angles, that the effective PRF is used.}.  
\item[Regress] the $N$ convolved layers with the real map of the sky, ideally weighted by the noise\footnote[7]{As dictated by the formalism of \S~\ref{sec:formalism}, take care that the mean of the pixels to be fit in each layer equals zero.}.
\end{description}

Stacking should be performed on maps in $\rm Jy\, beam^{-1}$. 
Errors can be estimated with a bootstrap technique,  
as described in \S~\ref{sec:errors}.  
Systematic errors in the method include beam area and calibration uncertainties.  Note that calibration errors may be correlated between bands of the same instrument---an effect that should be accounted for when fitting models to stacked flux densities.    
%#######################
\subsection{Testing the Method}
\label{sec:simulations}
% #######################
 %%%%%%%%%%%%%%%%%%%%
\begin{figure}[!t]
\centering
\vspace{-2.6mm}
\hspace{-5mm}
\includegraphics[width=0.5\textwidth]{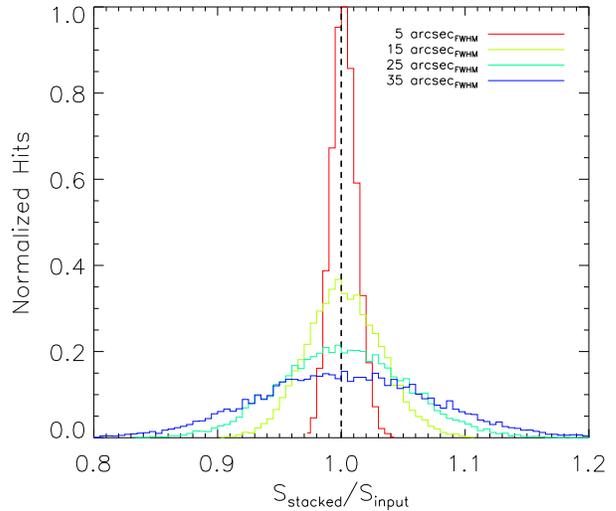}
\caption{Test of the traditional stacking estimator on simulated maps with randomly distributed (i.e., unclustered) sources.  Histograms show the resulting output vs.\@ input flux densities of 10{,}000 iterations per beam size, for beams ranging from $\rm FWHM= 15$--35\arcsec,  and a source density  of $\sim 2\, \rm arcmin^{-2}$.  The vertical dashed line at unity represents an unbiased estimate.  For all beam sizes, the estimator is shown to be unbiased, though the errors increase with an increased number of sources per beam (i.e., for larger beams).   }
\label{fig:hist}
\end{figure}
%%%%%%%%%%%%%%%%%%%%
 %%%%%%%%%%%%%%%%%%%%
\begin{figure}[!t]
\centering
\vspace{-2.6mm}
\hspace{-5mm}
\includegraphics[width=0.5\textwidth]{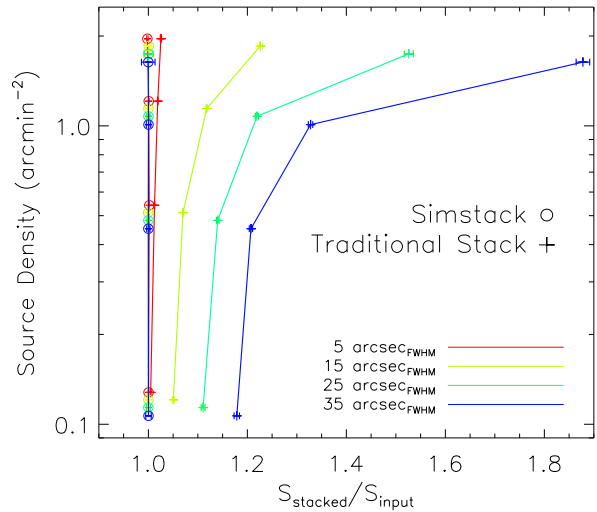}
\caption{Test of the traditional and simultaneous stacking estimators on 10{,}000 clustered simulated maps.  Recovered vs.\@ input fluxes are measured as a function of source density and beam size.  The traditional estimator performs well for beams smaller than 5\,arcsec, but quickly becomes  biased for bigger beams, particularly at higher source densities.  The simultaneous stacking algorithm, \simstack, on the other hand returns an unbiased estimate in all cases.
}
\label{fig:bias}
\end{figure}
%%%%%%%%%%%%%%%%%%%%
%%%STACKED FLUX FIGURE%%%%%
\begin{figure*}%[t!]
\centering
%\vspace{-2.6mm}
\hspace{-6mm}
\includegraphics[width=1.025\textwidth]{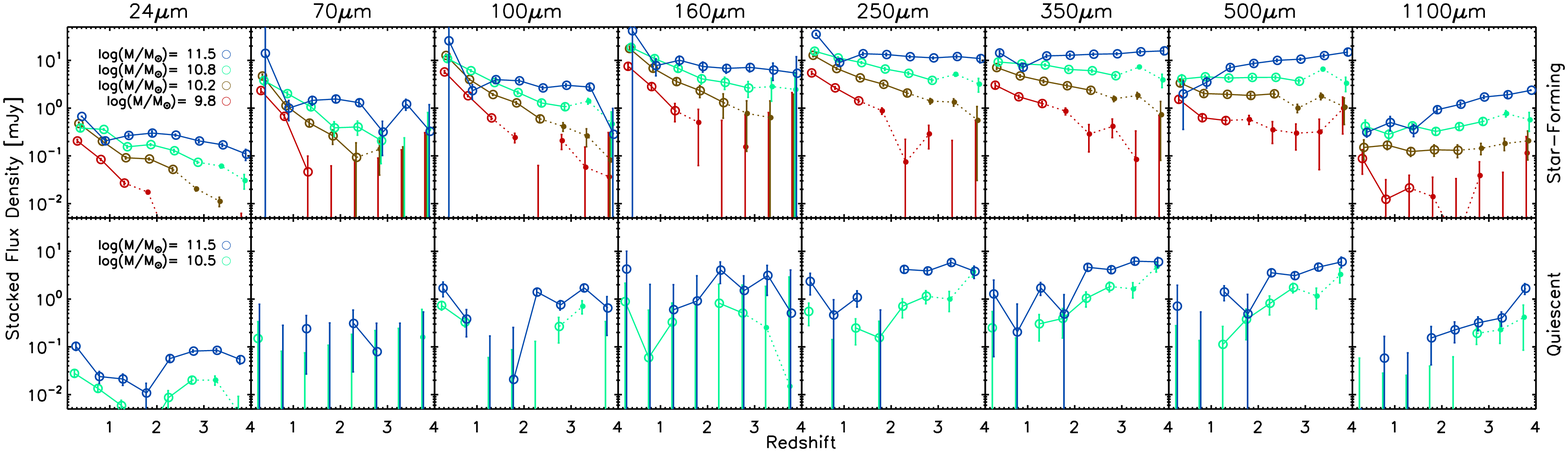}
%\vspace{-5mm}
\caption{Stacked flux densities vs.\@ redshift for star-forming  (top row) and  quiescent  (bottom row) galaxies, in divisions of mass.  
Open circles represent bins with greater than 50\% completeness.  Note that the flux densities shown here have been  color-corrected (see \S~\ref{sec:color_correction}).  Data and errors are tabulated in Tables~\ref{tab:stacked_sf} and \ref{tab:stacked_passive} for star-forming and quiescent galaxies, respectively.  
}
\label{fig:stacked_fluxes}
\end{figure*}
%%%%%%%%%%%%%%%%%%%%
Monte Carlo simulations consisting of 10{,}000 iterations are performed to test the estimator for biases.
Two sets of simulated maps---one containing Poisson distributed (random) sources and the other of realistically correlated (clustered) sources---are constructed.  
Each map is a superposition of sources of varying mass, with the number of sources in each mass bin the same as that of real data.
The flux densities given to the sources are drawn from Gaussian distributions centered on the flux densities of the measured mean stacked flux densities, and the width of the Gaussian five times that of the uncertainty on the stacked values, in order to introduce a significant level of stochasticity.

Sources in clustered simulated maps have their positions drawn from the actual positions of the catalog sources, in order to properly reproduce higher order correlations. 
Each map is then convolved with a Gaussian kernal approximating the instrumental beams, with FWHM values ranging from 5 to 35\,arcsec.

We then perform stacking analyses on the maps in two ways: 1) with a traditional \citep[e.g.,][]{marsden2009} estimator; and 2) with the \simstack\  
algorithm.  Finally, the stacked flux densities are compared to the known input mean values.  The histogram in Figure~\ref{fig:hist} illustrates how the traditional estimator returns an unbiased estimate of the mean flux ($S_{\rm stacked}/S_{\rm input} \approx 1$) of an unclustered simulation, but with an uncertainty that depends on the beam size  \citep[see also][]{viero2012}.   Similarly, Figure~\ref{fig:bias} shows the bias vs.\@ source density (in number of sources per square arcmin) for clustered simulated maps, where the traditional estimator is represented by crosses and our simultaneous stacking method by circles.  The data points are offset vertically for visual clarity.  The traditional stacking estimator is relatively faithful for beams of $\rm FWHM \le 5\arcsec$, but quickly becomes biased, especially in catalogs with many sources.  The simultaneous stacking instead returns an unbiased estimate of the mean flux density, with errors $\sigma = 1$--3\%, increasing with increasing beam size.
%\newpage
%#######################
\subsection{Estimating Uncertainties}
\label{sec:errors}
% #######################
In addition to measurement errors, we must account for potential systematic errors introduced by photometric redshifts.  
To address this potential bias we developed an extension of the typical bootstrap technique, hereafter referred to as the extended bootstrap technique, or EBT.  
Like a typical bootstrap, the EBT assembles new bins for stacking from the sources in the parent catalog, 
with the difference that rather than simply drawing sources randomly from the original bin, \emph{all} of the sources in the catalog are first
perturbed according to their redshift uncertainties, and then new bins are assembled from the new redshifts and masses.  
Simulated redshifts are determined by drawing randomly from the redshift probability distribution output by the photometric redshift code, {\sc EASY}.   
Simulated masses, which change depending on redshift of the source, must be estimated as well.  
However, as estimating a new mass for every new redshift for every source would be overly labor intensive---particularly considering that most perturbations from the nominal redshift are rather small---we instead use the fact that the mass is a strong function of  \emph{K}-band magnitude with some slight additional dependence on $J-K$ color, and we estimate the new mass using the perturbed redshift and the observed magnitude and color.  
Finally, each simulated catalog is split up into bins resembling those of the original stack and new stacked flux densities are estimated with \simstack.  This is done  1{,}000 times.  
Measurement and bootstrap errors are then added in quadrature, though we note that the error budget is dominated by the EBT estimates. 

The EBT accounts for the possibility of cross-contamination of galaxies across redshift and mass bins, in addition to the stochasticity of the catalog members measured by the traditional bootstrap technique, thus resulting in a more realistic error.   
We find that the EBT increases uncertainties by an average of 22\%; where the correction in bins with better photometry is less; while the correction in bins with poor photometry (i.e., high redshift and/or low mass) can be as much as 50\%.  Note that these uncertainties account for both instrumental and confusion noise, as well as for any pixel-pixel correlations that map-making may introduce.  

Lastly, systematic errors arising from estimating the solid angles (or beam areas) of the MIPS PSFs \citep[][]{bethermin2010}, as well as calibration uncertainties at all wavelengths \citep[][]{engelbracht2007,gordon2007,stansberry2007,swinyard2010}, 
must be taken into account, particularly when estimating the contribution to the CIB from galaxies.  
These errors are accounted for empirically through inclusion into the Monte Carlo simulation used to estimate the ultimate errors.    
   
\newpage
% ======================
\section{Results}
\label{sec:results}
% ======================

%#######################
\subsection{Stacked Flux Densities}
\label{sec:stacked_flux}
% #######################
Stacked flux densities and 1$\sigma$ uncertainties are shown for star-forming and  quiescent galaxies in the top and bottom panels of Figure~\ref{fig:stacked_fluxes}, and listed in Tables~\ref{tab:stacked_sf} and \ref{tab:stacked_passive}, respectively.   
We find statistically significant signals in the majority of the  bands and bins, with the noisiest signals from bins of lowest masses and highest redshifts, and the most robust signals in those of the higher mass bins.  Also, as expected from the beam size and noise properties of the maps, the MIPS 24\rmicron\ and three SPIRE bands return stacked flux densities with the highest signal-to-noise, while the 70 and 160\rmicron\ uncertainties are larger.  
The uncertainties at 1100\rmicron\ are also higher, but that is largely a reflection of the area of the AzTEC field, which is half that of the other bands.  

We note that the traditional stacking method, as anticipated from simulations (\S~\ref{sec:simulations}), returns systematically higher results, with the bias proportional to the strength of the  clustering, which increases with increasing stellar mass.  Considering this trend, any method that applies one correction for all stacked results should be viewed with suspicion.   

Also notable is the significant contribution from galaxies identified as  quiescent by their colors, a signal which is most prominent from the galaxies in the highest redshift bins.  Their flux densities in all bands increase steadily with increasing redshift, to the point where at $z\gsim 3$, they are comparable to those of the most massive star-forming galaxies in the sample.  It is likely that this is the result of  misclassification of star-forming galaxies arising from low signal-to-noise photometry scattering their colors into the quiescent plane of the \emph{UJV} diagram.  We discuss this and other scenarios in \S~\ref{sec:quiescent}.

%#######################
\subsection{Best-Fit SEDs}
\label{sec:best_fit_sed}
% #######################

\begin{figure*}%[h!]
\centering
%\vspace{-2.6mm}
\hspace{-13.3mm}
\includegraphics[width=1.07\textwidth]{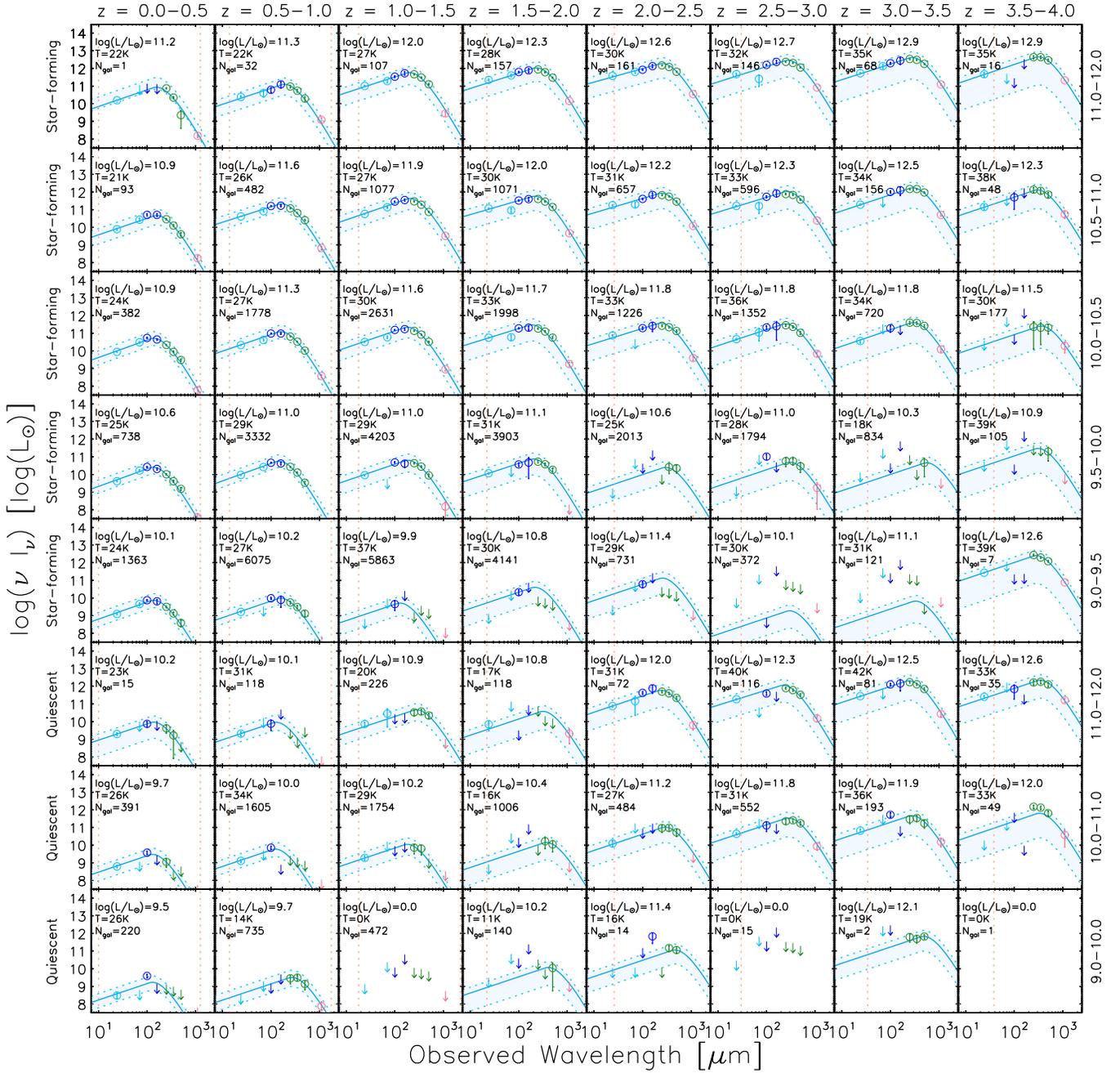}
\caption{Stacked intensities ($\nu I_{\nu}$) of mass-selected sources in the {\sc UDS} field divided into bins of mass and redshift.
Data at 24 and 70\rmicron\  (light blue circles) are from \emph{Spitzer}/MIPS; 
those at 100 and 160\rmicron\ (dark blue circles) are from \emph{Herschel}/PACS; 
those at 250, 350, and 500\rmicron\ (green circles) are from \emph{Herschel}/SPIRE; and those at 1100\rmicron\ (red circles) are from AzTEC. 
The error bars represent the 1$\sigma$ Gaussian uncertainties estimated with the extended bootstrap method described in \S~\ref{sec:errors}.
Non-detections are shown as 2$\sigma$ upper limits plotted as downward pointing arrows.   
The median values of the redshift distributions are used to convert flux densities into rest-frame luminosities.
The SED is modeled as a modified blackbody with a fixed emissivity index, $\beta = 2.0$, and a power-law approximation on the Wien side with slope $\alpha =1.9$.  The solid blue line in each panel is the best-fit SED, and the shaded region enclosed by dotted light-blue lines shows the systematic uncertainties due to the width of the redshift distribution (interquartile range), which dominates the error budget.
}
\label{fig:sed}
\end{figure*}
%%%%%%%%%%%%%%%%%%%%
Intensities, $\nu I_{\nu}$, are estimated from stacked flux densities and plotted in Figure~\ref{fig:sed}    
with detections shown as circles, while non-detections are shown as 2$\sigma$ upper limits.  
Stacked flux densities trace out the SED 
of thermally emitting warmed dust.
While the shape of the SED is a superposition
of many blackbody emitters of different temperatures \citep[e.g.,][]{wiebe2009},
it has been shown to be well modeled 
as a modified blackbody of the form;
\begin{equation}
S_{\nu} = A \nu^{\beta} B(\nu,T),
\end{equation}
where $B(\nu,T)$ is the blackbody spectrum (or Planck function) with amplitude $A$,   
and $\beta$ is the emissivity index. 
The mid-infrared exponential on the Wien side of the spectrum can be replaced with a power-law of
the form $f_{\nu} \propto \nu^{-\alpha}$, which is added by specifically requiring that the two functions and their first
derivatives be equal at the transition frequency.
Values for both $\beta$ and $\alpha$ in the literature range from 1.5 to 2 \citep[e.g.,][]{blain2003,dunne2011,viero2012}, while
we use $\beta=2$ and $\alpha=1.9$, which represent the mean values of the best fits of the individual SEDs.   
Note that for both $\beta$ and  $\alpha$, we check that the exact values chosen does not significantly bias the result.

Our SED fitting procedure estimates the amplitude and temperature of the
above template.  For the SPIRE points, the SED fitting procedure \citep[described
in detail in][]{Chapin2008} takes the width and shape of the
photometric bands into account, as well as the absolute
photometric calibration uncertainty in each band. 
Correlations due to instrumental noise
are estimated and accounted for with a Monte Carlo procedure.

Correlated confusion noise must also be accounted for in the fit,
as these correlations reduce the significance of 
a detection in single band.  
That is, not accounting for correlated noise in the measurements of, say, the three SPIRE bands,  
would lead to attributing additional weight in the overall fit to these data, potentially leading to a bias. 
This is discussed in more detail in \citet{Moncelsi2011} and \citet{viero2012}.      
We estimate the Pearson coefficients of
the correlation matrix for all bands from the beam-convolved maps (Table~\ref{tab:covariance_matrix}). 
%%%%%%%%TABLE 2%%%%%%%%%
\input{table2.tex}
%%%%%%%%TABLE 2%%%%%%%%%

Interquartile errors, which reflect the uncertainty in dimming due to
the width of the redshift bins, are estimated from the distribution of redshifts over the full set of simulated catalogs generated as described in \S~\ref{sec:errors}.   
Best-fit SEDs for each mass and redshift bin are shown in Figure~\ref{fig:sed} as solid blue lines; dotted blue lines and shaded regions represent the  interquartile errors. 

The best-fit SEDs serve several purposes.  The first, and our primary purpose, is to infer the contribution from galaxies in each bin to the entire CIB (spanning the full range of wavelengths of our sample).   
Another purpose is to estimate infrared luminosities, as described in \citet[][]{kennicutt1998}, by integrating the SED between rest-frame 8 and 1000\rmicron\ (shown as horizontal yellow dotted lines in panels of Figure~\ref{fig:sed}).  These are later used when quantifying the contribution to the CIB from galaxies classified as \lq\lq normal\rq\rq, luminous, and ultra-luminous infrared galaxies (\S~\ref{sec:luminosity}).  Infrared luminosities can be used as an indicator of obscured star formation, a topic that will be explored in Arumugam et al.\@ (in prep.).  
Finally, best-fit SEDs give a measure of the effective dust temperatures, which we discuss in \S~\ref{sec:temperature}.  For reference, both temperatures and luminosities are listed in the panels of Figure~\ref{fig:sed}.   

%#######################
\subsection{Total Resolved CIB}
\label{sec:resolved_cib}
% #######################

We estimate the contribution  to the CIB from our \emph{K}-selected galaxy sample (first without correcting for incompleteness) 
by multiplying the emission from each bin ($\nu I_{\nu}$) by the number of galaxies in that bin, and summing them together.  
Results are tabulated in the second column of Table~\ref{tab:cib_totals}, labeled \lq Total Stacking\rq.    

Next, corrections for incompleteness  are made for samples that are more than 50\% complete, 
by dividing each bin by its completeness estimate (drawn from Figure~\ref{fig:comp}) before summing.   
Results are tabulated in the third column of Table~\ref{tab:cib_totals}, labeled \lq Completeness Corrected\rq.  
The choice of 50\%, though somewhat arbitrary, is chosen because at that point the uncertainty in the incompleteness estimate is not yet greater than the correction.   Note that the uncertainties associated with this correction are accounted for as part of the Monte Carlo simulation used to estimate the total uncertainties.   
We find that the completeness correction adds between $\sim 1$--3\% to the total CIB. 
If we relax the completeness requirement and correct bins which are as little as 10\% complete, the completeness correction rises to $\sim 8$--15\% and the resolved CIB to 97\% in the SPIRE bands.  
Although hardly robust, this at least suggests that some fraction of the remaining CIB originates from faint sources. %; 
This scenario is discussed in more detail in \S~\ref{sec:faint_sources}.   
In all subsequent CIB figures, plotted points are completeness-corrected unless otherwise noted, with the total contribution in each band plotted as pink squares.  

Also plotted are estimates of the total CIB as measured by: 
\emph{Spitzer}/MIPS at 24\rmicron\  \citep[diamond;][]{dole2006};  
\emph{IRAS} at 60\rmicron\ \citep[boxes;][]{miv2002a};  
\emph{Spitzer}/MIPS at 24, 70, and 160\rmicron\ \citep[asterisks;][]{bethermin2010} as well as at  70 and 160\rmicron\ \citep[exes;][]{jauzac2011}; 
 WHAM at 100, 140, and 240\rmicron\ \citep[crosses;][]{lagache2000}; 
 and from \emph{COBE}/FIRAS  spectra spanning $\sim 200$ to 1200\rmicron\ \citep[solid line;][]{lagache2000}.
Lower limits are shown as upward pointing arrows from \emph{Spitzer}/MIPS at 24\rmicron\  \citep{papovich2004} and at 70 and 160\rmicron\ \citep{dole2006};  
SPIRE at 250, 350, and 500\rmicron\ \citep{bethermin2012b}; 
and SCUBA at 450\rmicron\ \citep{serjeant2004} and 850\rmicron\ \citep{smail2002}.   
Lastly, plotted as lavender asterisks is the resolved CIB using 24\rmicron\ priors from Vieira et al.\@ (in prep.)

As previously described, the relationship between stacked fluxes from different bands is well approximated by a simple thermal dust SED. 
This can be used to roughly estimate the total resolved CIB between bands, as well as give us a better handle on the contribution from noisy bands. 
Thus, we estimate the total contribution to the CIB from summing the best-fit SEDs of Figure~\ref{fig:sed} (corrected for incompleteness), and report them in the fourth column of Table~\ref{tab:cib_totals}, labeled \lq Total Model SEDs\rq. 
Comparing these measurements to the absolute CIB values listed in the last column of the same Table, labeled \lq Reference\rq, we find that our full sample resolves 
$(69\pm 15)\%$, $(78\pm 18)\%$, $(58\pm 13)\%$, $(78\pm 18)\%$, $(80\pm 17)\%$, $(69 \pm 14)\%$, 
$(65\pm 12)\%$, and $(45\pm 8)\%\,\rm nW\,m^{-2}\,sr^{-1}$ at 
24, 70, 100, 160, 250, 350, 500, and 1100\rmicron, respectively.   

%%%%%%%%%%%%%%%%%%%%%%%%%%%%%%%%%%%%%%%%
\begin{figure}[t!]
\centering
\vspace{-2.6mm}
%\hspace{-4mm}
\includegraphics[width=.50\textwidth]{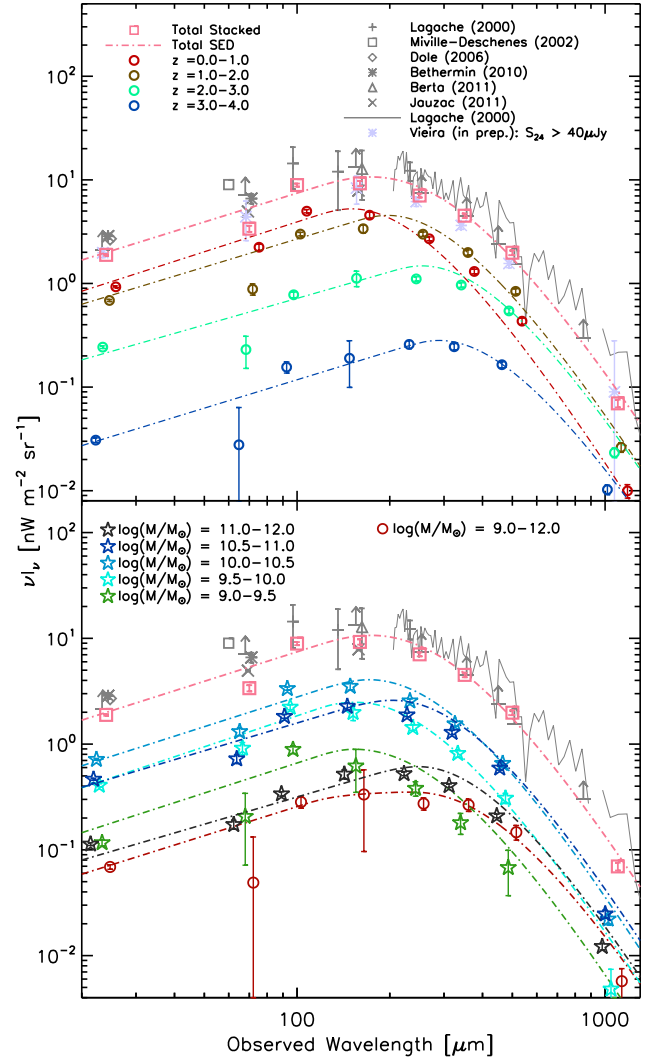}
\vspace{-2.6mm}
\caption{
{\bf Top Panel:}~~Contribution to the CIB from equally spaced redshift slices.  Measured data are shown as circles with error bars, while best-fit SEDs are shown at dot-dashed lines.  
Also plotted as lavender asterisks is the resolved CIB using 24\rmicron\ priors from Vieira et al.\@ (in prep.).  
{\bf Bottom Panel:}~~Contribution to the CIB in divisions of mass, with star-forming galaxies shown as stars and quiescent galaxies (with all mass bins combined) shown as circles.  
{\bf Both Panels:}~~ Also plotted is a selection of measurements of the total CIB in grey, from: 
\emph{Spitzer}/MIPS at 24\rmicron\  \citep[diamond;][]{dole2006};  
\emph{IRAS} at 60\rmicron\ \citep[boxes;][]{miv2002a};   
\emph{Spitzer}/MIPS at 24, 70 and 160\rmicron\ \citep[asterisks;][]{bethermin2010};  
and 70 and 160\rmicron\ \citep[exes;][]{jauzac2011};  
\emph{Herschel}/PACS at 160\rmicron\ \citep[triangles;][]{berta2011};   
WHAM at 100, 140, and 240\rmicron\ \citep[crosses;][]{lagache2000}; 
and \emph{COBE}/FIRAS  spectra from $\sim 200$ to 1200\rmicron\ \citep[solid line;][]{lagache2000}.  
Lower limits are from \emph{Spitzer}/MIPS at 24, 70, and 160 \rmicron\  \citep{papovich2004,dole2006}; 
SPIRE at 250, 350, and 500\rmicron\ \citep{bethermin2012b};  
and SCUBA at 450 and 850\rmicron\ \citep{smail2002,serjeant2004}.    
Our total CIB measurements and best-fit SEDs are shown in both panels as pink boxes and dot-dashed lines, respectively. 
}
\label{fig:bethcib}
\end{figure}
%%%%%%%%%%%%%%%%%%%%

%%%%%%%%TABLE 3%%%%%%%%%
\input{table3.tex}
%%%%%%%%TABLE 3%%%%%%%%%

\subsection{Contribution to the CIB in Broad Redshift Bins}
\label{sec:z}
% #######################
Plotted as open circles in the top panel of Figure~\ref{fig:bethcib} and tabulated in Table~\ref{tab:cib_redshifts} are estimates of the contribution to the total resolved CIB in four redshift bins.  
The dot-dashed lines connecting the circles represent the equivalent summed SED fits.  
Notable is the striking dependence on the contribution to different bands from different redshifts---a result of negative $K$-correction---with the  $z=0$--1 bin dominating the CIB at $\lambda \lsim 160$\rmicron, and the  $z=1$--2 bin the chief contributor at $\lambda \gsim 250$\rmicron.  

Not surprisingly, as infrared luminosity is a tracer of obscured star formation \citep[e.g.,][]{kennicutt1998}, this behavior closely mimics the rapid rise in the star-formation history of the universe \citep[e.g.,][]{hopkins2006,behroozi2013a},  peaking at $z\sim 1$, 
as well as the same general trends as the model predictions of \citet[][top panel of figure~11]{bethermin2011}, though it should be noted that the limits of the redshift bins are not identical.  
We explore similarities with models and agreement with measurements of the infrared luminosity density further in \S~\ref{sec:luminosity}.  
%\newpage  
%#######################
\subsection{Contribution to the CIB in Stellar-Mass Bins}
\label{sec:mass}
% #######################
We estimate the contribution to the CIB in divisions of stellar mass by summing completeness-corrected emission in rows of Figure~\ref{fig:sed}, with star-forming and  quiescent galaxies grouped separately, and tabulated in Tables~\ref{tab:cib_masses_sf} and \ref{tab:cib_masses_passive}, respectively.  They are also shown in the bottom panel of Figure~\ref{fig:bethcib} as color-coded stars and circles (quiescent galaxies are all grouped together), with the dot-dashed lines representing the equivalent summed SED fits.  

A significant part of the resolved CIB ($\gsim 65\%$) at all wavelengths appears to originate from star-forming galaxies having stellar masses between log($M/\rm M_{\odot}) = 10.0$--11.0.  
The total contribution from more massive galaxies is $\sim 5$\%, while that from less massive galaxies is 30\%.  
Galaxies from the log($M/\rm M_{\odot}) = 10.0$--10.5  bin provide the highest contribution everywhere; while those from the log($M/\rm M_{\odot}) = 10.5$--11.0 bin provide second highest contribution at $\lambda \gsim160$\rmicron, and those from the log($M/\rm M_{\odot}) = 9.5$--10.0 bin provide the second highest contribution at shorter wavelengths.  
Quiescent galaxies together, unsurprisingly, contribute very little to the total resolved background, of order 5\%.   

Galaxies form in dark matter over-densities, or halos \citep[e.g.,][]{mo1996}, with the peak efficiency for star formation in halos of log($M/\rm M_{\odot}) \sim 12.0$ \citep[e.g.,][]{moster2010,behroozi2010,bethermin2012a, viero2013a}, which appears to be remarkably consistent throughout the age of the Universe \citep{behroozi2013}.  
On the other hand, the stellar mass of the galaxies which formed in these halos, and the evolution of that relationship, is less certain.    
\citet{wang2013} found stellar-to-halo mass values of $\sim 10^{-2}$ to $10^{-3}$ from to $z=0$ to 2, which equates to log($M/\rm M_{\odot}) \sim 10.4$ to 10.6, largely consistent with our findings.  
We explore the luminosity density evolution of these same galaxies in the next section.

%#######################
\subsection{Contribution to the CIB as a Function of Galaxy Luminosity}
\label{sec:luminosity}
\begin{figure}[t!]
\centering
%\vspace{-5mm}
\hspace{-7mm}
\includegraphics[width=0.5\textwidth]{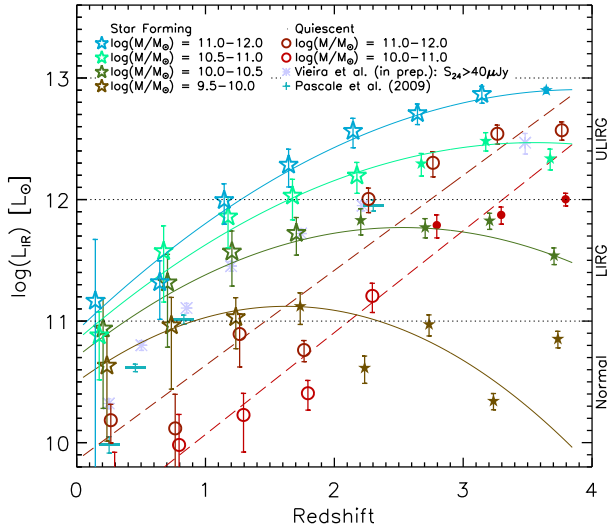}
\caption{
Luminosity vs.\@ redshift in divisions of mass for star-forming (stars) and quiescent galaxies (circles).  
Open symbols represent bins with greater than 50\% completeness.  
Polynomial fits to the data are plotted for star-forming (solid lines) and quiescent (dashed lines) galaxies. 
Notable is the rapid evolution in luminosity of the quiescent population, possibly due to noise in the optical photometry of higher redshift sources scattering star-forming galaxies into the quiescent section of the \emph{UVJ} plane.   
Despite this enhanced luminosity, galaxies identified as quiescent provide only about 5\% to the total CIB (\S~\ref{sec:mass}).  
Results from stacking of 24\rmicron-selected sources from \citet{pascale2009} and Vieira et al.\@ (in prep.)\@ are plotted as cyan crosses and lavender asterisks, respectively. } 
\label{fig:lir}
\end{figure}

% #######################
Infrared luminosities, $L_{\rm FIR}$, are calculated by integrating 
the rest-frame SEDs between 8 and 1000\rmicron\ \citep[][]{kennicutt1998}.  
In Figure~\ref{fig:lir} we plot  $L_{\rm FIR}$ as a function of redshift, in divisions of stellar mass, with star-forming and quiescent galaxies displayed as stars and circles, respectively, and open symbols represent bins whose completeness is greater than 50\%.     
Infrared galaxies have conventionally been classified by their luminosities into \lq\lq normal\rq\rq\ ($ L < 10^{11}\,\rm L_{\odot}$), luminous (LIRG: $ L = 10^{11-12}\,\rm L_{\odot}$) and ultra-luminous (ULIRG: $ L = 10^{12-13}\,\rm L_{\odot}$) infrared galaxy classes, illustrated in Figure~\ref{fig:lir} as horizontal dotted lines and right-handed labels.  

We fit simple polynomials to $L(M_i,z)$ vs.\@ \emph{z} to each stellar-mass bin, $i$, such that 
\begin{equation}
 {\rm log}(L(M_i,z))=\sum_{j=0}^{n} x_{i,j}  z^j, 
\end{equation}
following the rule that the \emph{i} variables of each $x_{i,n}$ must additionally obey their own polynomial fit,  
\begin{equation}
x_{i,j}=\sum_{k=0}^{n} y_{j,k} {\rm log}(M^k_i),    
\end{equation}
where the order of the polynomial fit, $n$, is 2 and 1 for star-forming and quiescent galaxies, respectively. 
This simple parameterization ties together the evolution of the luminosity with stellar mass and redshift, and allows us to fully explore the \emph{L--M--z} space\footnote[8]{The fitting functions to produce these curves are made available online.}.      
We find  

\begin{equation}
{\bf y_{\rm star\,forming}} = 
\begin{pmatrix}
-7.248 & 3.160 & -0.137 \\
-1.634 & 0.335 & -0.009 \\
-7.758 & 1.374 & -0.062 \\
\end{pmatrix}
, 
\end{equation}
\begin{equation}
{\bf y_{\rm quiescent}} = 
\begin{pmatrix}
2.672 & 0.624  \\
1.430 & -0.056  \\
\end{pmatrix}, 
\end{equation}
and plot them respectively as solid and dashed lines in Figure~\ref{fig:lir}.

We find a rapid rise of luminosity with redshift for the most massive populations, and an apparent turnover at $z\gsim 2$ in the less massive ones; though we caution that incompleteness makes this turnover effect difficult to interpret.   
Particularly striking is the evolution of the luminosities of quiescent galaxies, since at lower than $z\sim 2$ they are barely detectable, while by $z\sim 3$ they are nearly as luminous as star-forming galaxies of similar mass.   This is likely partially due to misclassification of star-forming galaxies arising from low signal-to-noise photometry at high redshift scattering galaxies into the quiescent part of the \emph{UVJ} plane. 
Note that although the most massive galaxies at high redshift are very luminous, they make up a relatively small fraction of the full catalog and thus contribute only $\sim 4\%$ of the total resolved CIB (see \S~\ref{sec:mass}, and bottom panel of Figure~\ref{fig:bethcib}).  
It is possible that some fraction of the dust heating is due to active galactic nuclei (AGN), as similar behavior has been seen in individually resolved objects at 24\rmicron\ \citep[e.g.,][]{perez2008,marchesini2010}, though the optical SEDs of this population on average are not indicative of AGN.  
These scenarios are discussed in more detail in \S~\ref{sec:quiescent} and \S~\ref{sec:agn}.

Also shown are 24\rmicron-selected stacking results from \citep[][cyan crosses]{pascale2009} and Vieira et al.\@  (in prep., lavender asterisks).  Their selection groups together galaxies of all masses and colors making a direct comparison difficult, yet the general trend of both stacks seem to agree reasonably well.  
 
%%%%%%%%%%%%%%%%%%%%%%%%%%%%%%%%%%%%%%%%
%%%%%%%%%%%%%%%%%%%%

 %LUMINOSITY DENSITY FIGURE
%%%%%%%%%%%%%%%%%%%%%%%%%%%%%%%%%%%%%%%%
\begin{figure}[t!]
\centering
\hspace{-13mm}
\includegraphics[width=0.5\textwidth]{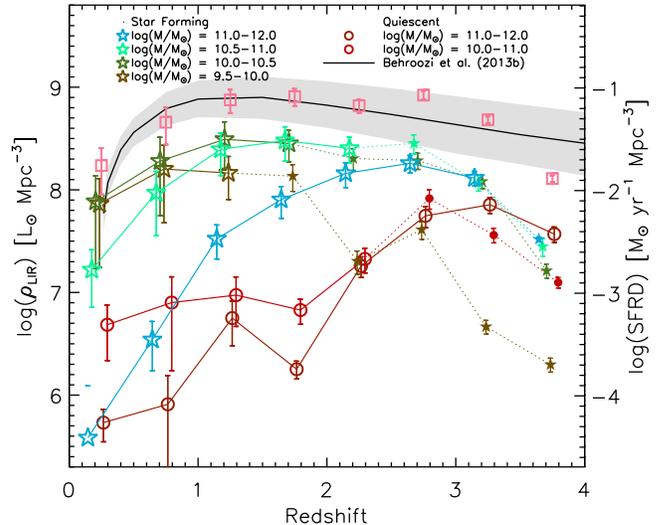}
\caption{
Luminosity density vs.\@ redshift in divisions of mass for star-forming (stars) and quiescent galaxies (circles).  
Open symbols represent bins with greater than 50\% completeness.  
We find a rapid rise in the contribution of the most massive galaxies to the luminosity density, as well as a steep decline in less massive galaxies, indicative of downsizing.    
Pink squares are totals for each redshift, which agree well with the model of \citet[][]{behroozi2013a}.  
} 
\vspace{3mm}
\label{fig:lird}
\end{figure}
%%%%%%%%%%%%%%%%%%%%
 %LUMINOSITY DENSITY
In Figure~\ref{fig:lird} we explore the contribution to the infrared luminosity density \citep[$\rho_{\rm L_{IR}}$; e.g.,][]{hopkins2006,lefloch2005,rodighiero2010,behroozi2013a} in divisions of stellar mass and color.  
Again, star-forming and quiescent galaxies are displayed as stars and circles, respectively, and open symbols represent bins with greater than 50\% completeness.  
We find that the contribution from the most massive galaxies (which we have already found account for less than 4\% to the total CIB) evolves rapidly with redshift, such that by $z > 2$ they are responsible for as much as their more abundant, less massive counterparts---a clear demonstration of downsizing \citep[e.g.,][]{cimatti2006,fontanot2009}.   

Plotted as pink squares is the total $\rho_{\rm L_{IR}}$, as well as a model prediction for the star-formation rate density (SFRD) from  
\citet[][]{behroozi2013a}, converted to luminosity density using the \citet{kennicutt1998} relation 
\begin{equation}
SFR\,[{\rm M_{\odot}\, yr^{-1}}] = 1.728\times 10^{-10}\, L_{\rm IR}\, [\rm L_{\odot}],
\end{equation}
with an additional lowering by 0.23\,dex to convert to a \citet{chabrier2003} initial mass function \citep[e.g.,][]{kriek2009a}.  
Equivalent values for the SFRD are shown for reference on the right-hand axis of Figure~\ref{fig:lird}.  
We find relatively good agreement with the model, which is also in agreement with a host of different measurements \citep[e.g.,][]{rodighiero2010, perez2008b, lefloch2005,caputi2007,casey2012b,burgarella2013,magnelli2013} and models \citep[e.g.,][]{hopkins2006}.

\begin{figure}[t!]
\centering
\vspace{-1mm}
%\hspace{-11mm}
\hspace{-7mm}
\vspace{-3mm}
\includegraphics[width=0.50\textwidth]{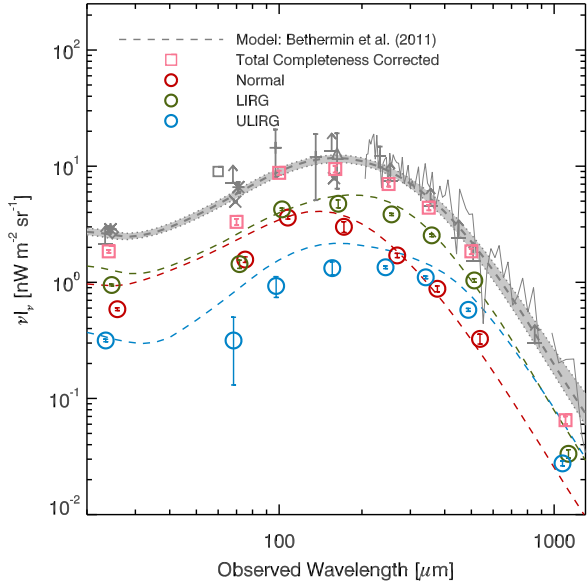}
\caption{Contribution to CIB from \lq\lq normal\rq\rq\ galaxies ($L < 10^{11}\,\rm L_{\odot}$), LIRGs  ($L < 10^{11-12}\,\rm L_{\odot}$), and ULIRGs  ($L < 10^{12-13}\,\rm L_{\odot}$).  
Normal galaxies and LIRGs contribute equally to make up most of the intensity at $\lambda \lsim 70$\rmicron, which is more sensitive to lower redshifts, while at longer wavelengths LIRGs and eventually ULIRGs contribute most to the signal.     
Also plotted are model predictions from \citet[][figure~13, bottom panel]{bethermin2010}, with the LIRG and ULIRG predictions somewhat high.
Although the model is a simple parametric fit to counts at multiple wavelengths, the high estimates for the LIRGs and ULIRGs lends weight to the suggestion that we are missing luminous, dust-obscured sources in our sample (\S~\ref{sec:missing_sources}).
} 
\label{fig:lcib}
\end{figure}

In Figure~\ref{fig:lcib} we explore the contribution to the CIB from  \lq\lq normal\rq\rq\ galaxies, LIRGs, and ULIRGs.   
Data are constructed by summing intensities of bins corresponding to their luminosities as determined from the best-fit SEDs (\S~\ref{sec:best_fit_sed}).   
Short of 160\rmicron,  the contribution from LIRGs and less-luminous galaxies is comparable, while at wavelengths longer than 160\rmicron, LIRGs clearly dominate the resolved CIB.  
The contribution from log($L/\rm  L_{\odot}) < 11$ galaxies falls rapidly at wavelengths greater than 160\rmicron, which may suggest a diminishing contribution from fainter populations at high redshift---which is again suggestive of downsizing---but it could also mean that fainter galaxies are simply being missed.  
The contribution from ULIRGs, which, as seen from Figure~\ref{fig:lir}, are located at $z\gsim 1$, peaks at longer wavelengths, and is an order of magnitude lower than less-luminous galaxies at $\lambda \lsim 160$\rmicron.    
Note that if small numbers of exceptionally luminous sources, ultra-luminous or hyper-luminous infrared galaxies, have unusually high luminosities with respect to their stellar masses (i.e., high specific luminosities) this plot would fail to capture their distribution accurately.   

Also overlaid in this figure are predictions from \citet[][bottom panel of Figure~11]{bethermin2011}, a parametric backwards-evolution model fit to counts at multiple wavelengths.  The general trends are well reproduced, while in detail, ULIRGs fall short of model predictions. 
As we discuss \S~\ref{sec:missing_sources}, this  may be an indication that highly dust-obscured galaxies are missing from our optical/NIR-based, mass-selected catalog \citep[e.g.,][]{dey1999}.

%%%%%%%%%%%%%%%%%%%%%%%%%%%%%%%%%%%%%%%%
%%%%%%%%%%%%%%%%%%%%

%#######################
\subsection{Average Temperature Evolution for Star-Forming Galaxies}
\label{sec:temperature}
% #######################
%%%%%%%%%%%%%%%%%%%%
\begin{figure*}%[!t]
\centering
%\vspace{-4.0mm}
%\hspace{-10mm}
\includegraphics[width=1.0\textwidth]{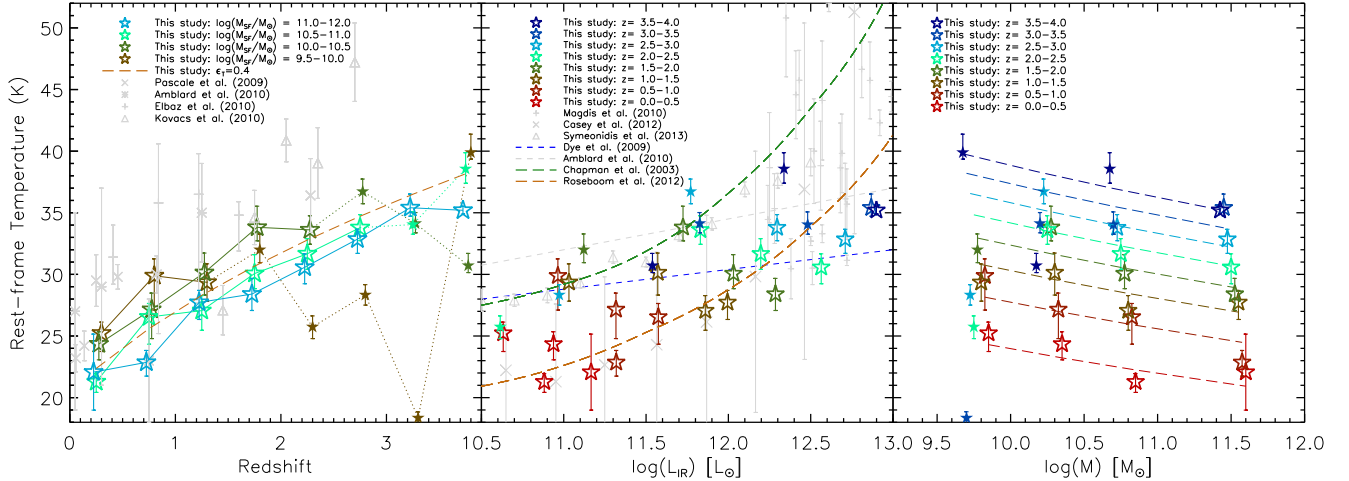}
\caption{
Average temperatures derived from the best-fit, modified blackbody SEDs vs.: redshift in the left panel; infrared luminosity in the center panel; and stellar mass in the right panel.   Open symbols represent bins with higher than 50\% completeness.  
{\bf Left Panel:}~ 
Temperatures of galaxies of all stellar masses are found to evolve strongly with redshift.  This evolution can be described as a power law with slope $\epsilon_{}=0.4\pm 0.1$ (orange dashed line).     
Also plotted are measurements from \citet[][pink exes]{pascale2009}; \citet[][asterisks]{amblard2010}; \citet[][crosses]{elbaz2010}, and  \citet[][triangles]{kovacs2010}.  
{\bf Center Panel:}~ 
The full ensemble of galaxy temperatures is shown to obey the canonical \emph{L--T} relation, described for local infrared galaxies by \citet[][green dashed line]{chapman2003} and 
at higher redshift by \citet[][red dashed line]{roseboom2012}.  
Also plotted are best-fits to BLAST and SPIRE sources from \citet[][grey dashed line]{amblard2010} and \citet[][blue dashed line]{dye2009}, respectively, and measurements from \citet[][crosses]{magdis2010}; \citet[][exes]{casey2012b}; and \citet[][triangles]{symeonidis2013}. 
{\bf Right Panel:}~ 
Conversely, the temperatures of galaxies appears to decrease with increasing stellar mass (and thus increasing $L_{\rm IR}$).  
Shown as dashed lines are tied power-law fits to the data at each redshift. } 
\vspace{3.0mm}
\label{fig:temp}
\end{figure*}
%%%%%%%%%%%%%%%%%%%%
In Figure~\ref{fig:temp} we plot temperatures derived from our best-fit SEDS as a function of redshift (left panel), infrared luminosity (center panel), and stellar mass (right panel) for star-forming galaxies divided into stellar-mass (left panel) or redshift bins (center and right panels).    
We emphasize that the reported temperatures are tied to the simple modified blackbody used to derive them (\S~\ref{sec:best_fit_sed}), and that if another model had been used \citep[e.g., a $\beta$ value of 1.5 instead of 2, a different opacity model, or a two component SED similar to that used by][]{dunne2001}, slightly different temperature values would have been derived \citep[also see][]{casey2012c}.   However, the \emph{trends} in the temperatures---either with redshift or with mass---should be relatively free of bias due to the model adopted.  Moreover, since our relatively high signal-to-noise measurements bracket the peak of the thermal SED, our ability to identify these trends is robust.  

We compare with temperature measurements of other galaxies---some FIR-selected, others NIR-selected---
noting that because our sample is mass-selected, we anticipate there to be discrepancies due to the selection functions of the different sets of galaxies.
In the leftmost panel we see a clear trend of an increase of temperature with redshift for galaxies of all stellar masses.  
This trend is also seen in stacked measurements from \citet[][pink exes]{pascale2009}, from submillimeter-selected galaxies from \citet[][asterisks]{amblard2010} and \citet[][crosses]{elbaz2010}, and \emph{Spitzer}-selected galaxies from  \citet[][triangles]{kovacs2010}.  The systematic offset of our mass-selected sample from these may partly be a result of the model fitted, and partly due to the fact that IR-selected sources will consequently be more luminous.  

Following \citet[][]{addison2013}, we fit the function 
\begin{equation}
T=T_0\left( \frac{1+z}{1+z_{\rm T}} \right)^{\epsilon_{\rm T}}, 
\end{equation}
where $T_0  = 27\, \rm K$ and $z_{\rm T}=1$ are the pivot temperature and redshift, respectively.  
This simple relationship is central to current halo models of the CIB \citep[e.g.,][]{addison2013,debernardis2012,shang2012,viero2013a}, yet remains poorly constrained, with values ranging from  $\epsilon_{\rm T} =  0.16 \pm 0.02$ \citep[][Model 2]{viero2013a} to $\epsilon_{\rm T} =  0.75\pm 0.10$ \citep{addison2013} .
We find $\epsilon_{\rm T} =  0.4\pm 0.1$, in good agreement with the CIB models presented in \citet[][Model 3]{viero2013a} and the \citet{lagache2013}. 

In the central panel we explore the well established Luminosity--Temperature (\emph{L--T}) relation \citep[e.g.,][]{dunne2000,dunne2011,dunne2001,dale2001,dale2002,chapman2003c,chapin2009,hwang2010,magdis2010,roseboom2012,magnelli2013},  comparing to relations measured by \citet[][green dashed line]{chapman2003},  \citet[][orange dashed line]{roseboom2012}, \citet[][blue dashed line]{dye2009}, and \citet[][grey dashed line]{amblard2010}, along with NIR-selected galaxies from \citet[][crosses]{magdis2010}, and FIR-selected galaxies from \citet[][exes]{casey2012b} and \citet[][triangles]{symeonidis2013}.   

We find poor agreement with the shallow slopes of \citet[][]{dye2009} and \citet[][]{amblard2010}, which may be a reflection of the shallow nature of the submillimeter data used in those studies (i.e., BLAST and H-ATLAS, respectively),  as well as the selection criteria (e.g., 3$\sigma$ detections in three SPIRE or PACS bands plus a 5$\sigma$ detection in SDSS or GAMA).  
On the other hand, we find generally good agreement with the trends reported locally in \citet{chapman2003} and at higher redshift by \citet{roseboom2012}; with overall values better described by the latter.  We also find that our measurements are consistent with the mean values of resolved, SPIRE-selected sources from \citet{casey2012b}, but that they are offset from those of \citet{symeonidis2013}.   

However, we notice that our incomplete bins (filled stars) have systematically higher temperatures, suggesting that incomplete samples in the optical/NIR select hotter sources.  This appears to be the case for the sources selected at NIR wavelengths by \citet{magdis2010}.  
This could also then mean that the selection criterion of \citet[][24\rmicron\ sources with 3$\sigma$ detections at 160\rmicron\ and either 100 or 250\rmicron]{symeonidis2013} favors sources with higher temperatures---in contrast to the missing \lq\lq hot dust\rq\rq\ ULIRG bias typically associated with submillimeter to millimeter sources \citep[e.g.,][]{eales2000,blain2004,chapman2004,chapman2008,casey2009,magdis2010,magnelli2010,hayward2012}.  
Again, the small discrepancies between these different measurements is likely attributable to a combination of selection effects and model-based systematic bias \citep[see also][]{magnelli2012}.     

Also of note in the central panel of Figure~\ref{fig:temp} is that in any one redshift bin the temperature appears to \emph{decrease} somewhat with increasing luminosity.  Similarly, in the rightmost panel we see that the temperature decreases for increasing stellar mass---particularly at lower redshifts---which follows since stellar mass and luminosity are strongly correlated (Figure~\ref{fig:lir}).  
We explore the evolution of temperature with stellar mass and redshift by fitting the relations with tied power-law fits, i.e.,
\begin{equation}
T(M,z)=A_{z} M_{z}^{\alpha_{{\rm T},z}},
\end{equation}
where power laws must obey the rule that the amplitudes and slopes of the fits are also fit by power laws
\begin{align}
A_{z}  = A_{z,0}+A_{z,1}(1+z)^{A_{z,2}}, \\
\alpha_{T}  =  \alpha_{T,0}+\alpha_{T,1}(1+z)^{\alpha_{T,2}}, 
\end{align}
%where ${\bf A} = [196,0.42,1.65]$ and ${\bf \alpha}=[5.77,-6.68,-0.01]$, 
where ${\bf A} = [-439.83,578.93,0.11]$ and ${\bf \alpha}=[-0.81,2.84\times 10^{-5},3.55]$, 
which amounts to steady increase of temperature with redshift, a mild anti-correlation of temperature with stellar mass, but a negligible change of the slope of the anti-correlation over time.  
 
There are several possible explanations for this apparent anti-correlation between stellar mass and temperature.  Dust creation is thought to be dominated by supernovae (SNe), and to a lesser extent, evolving main sequence stars on the asymptotic giant branch \citep[AGB; e.g.,][]{michalowski2010,michalowski2010b}.  
Since infrared luminosity \citep[an established proxy for star-formation rate; e.g., ][]{kennicutt1998} and dust mass \citep[e.g.,][]{santini2010,skibba2011} are strongly correlated with stellar mass, it follows that there would be more dust extended over a larger volume in more massive galaxies \citep[e.g.,][]{menendez2009}.  Also plausible is the possibility that less massive galaxies are more susceptible to stripping of their gas and dust \citep[e.g.,][]{abadi1999,mccarthy2008}, which \citet{rawle2012} show, can lead to higher temperatures.   

Thus, our findings appear to be in tension with the local (\emph{L--T}) relation for SMGs,  which may again be a function of the stellar-mass selection vs.\@ far-infrared selection.  Case in point:  \citet{hayward2013} show that galaxies with temperatures above roughly 40\,K are universally starbursts \citep[e.g.,][]{hernquist1989} as opposed to quiescently star forming \citep[e.g.,][]{dave2010}.  
As we already showed in \S~\ref{sec:luminosity} and Figure~\ref{fig:lcib}, there are indications that this is exactly the population that is being missed.  

%#######################
\subsection{Redshift Distribution of the Resolved CIB}
\label{sec:dsdz}
\begin{figure}[t!]
\centering
\vspace{-5mm}
%\hspace{-10mm}
%\vspace{5mm}
\includegraphics[width=0.50\textwidth]{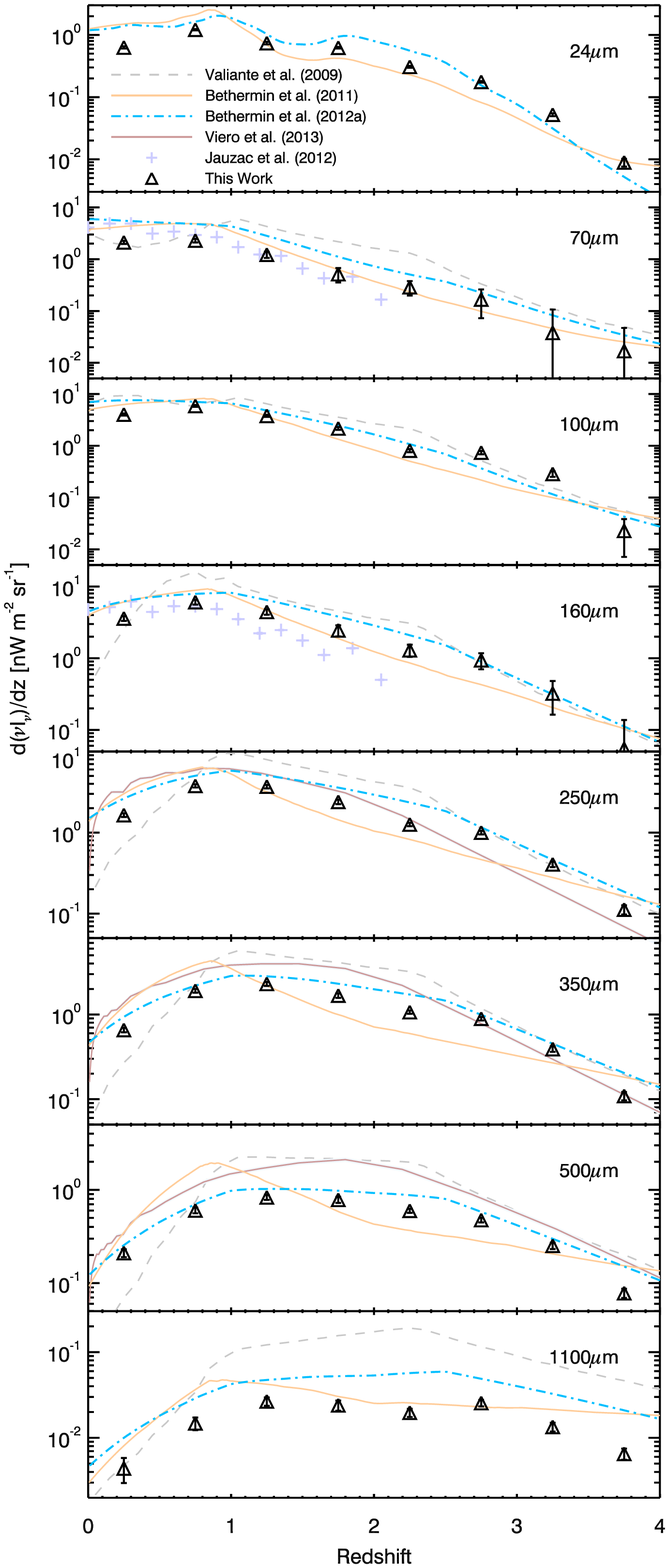}
\vspace{-10mm}
\caption{The redshift distribution in different bands of the resolved CIB emission, $d(\nu I_{\nu})/dz$, in roughly-spaced redshift slices.  Over plotted are models from \citet[][dashed gray]{valiante2009}; \citet[][solid yellow]{bethermin2011};  \citet[][dot-dashed blue]{bethermin2012c}; and \citet[][solid brown at 250, 350, and 500\rmicron]{viero2013a}; as well as published data from \citet[][lavender crosses at 70 and 160\rmicron]{jauzac2011}. %; MORE?? 
The \citet{bethermin2012c} model, which is determined from stacking 24\rmicron\ priors, fits remarkably well, suggesting a strong correlation between our different sets of catalogs.  
}
\label{fig:dsdz}
\end{figure}

% #######################
The redshift distribution of the resolved CIB emission, $d(\nu I_{\nu})/dz$, is measured by summing the completeness-corrected intensities of all stellar-mass bins separately in each band, and is plotted Figure~\ref{fig:dsdz} as upward pointing triangles, signifying that they are lower limits.  
The peak intensities shift from $z \sim 0.5$--2 with increasing wavelength, indicative of the peak of star formation occurring at $z\sim 1$--2 \citep[e.g.,][]{hopkins2006, behroozi2013a,wang2013},  
and the redshift sensitivity of the different bands.  

Also plotted are predictions for the entire CIB from a selection of recently published models. 
The \citet{bethermin2012c} model in particular, which is based in part on stacking measurements with 24\rmicron\ priors \citep{bethermin2012b}, describes the measured distribution extremely well.  
However, we caution that since the completeness is a strong function of redshift (see Figure~\ref{fig:comp}), it is not expected that the remaining CIB be distributed evenly in redshift.  

This measurement has multiple applications.  For example, CIB anisotropy measurements \citep[e.g.,][]{lagache2007,viero2009,hall2010,viero2013a,amblard2011,lagache2011,lagache2013} are typically interpreted with halo models \citep[e.g.,][]{peacock2000a, cooray2002}, which assign intensities to dark matter halos a function of halo mass, redshift, and SED of thermal dust emission.  
 These models find significant degeneracies between SEDs and the redshift distribution of the CIB \citep[e.g.,][]{addison2012,shang2012}, or halo bias and redshift distribution of the CIB \citep[e.g.,][]{holder2013}, limiting their interpretive power.  The measurements of $d(\nu I_{\nu})/dz$ reported here, as well as the \emph{T}-\emph{M}-\emph{z} dependence of the thermal SED from \S~\ref{sec:temperature},  provide powerful constraints for future models.   

%%%%%%%%%%%%%%%%%%%%%%%%%%%%%%%%%%%%%%%%
%%%%%%%%%%%%%%%%%%%%
% ======================
\section{Discussion}
\label{sec:discussion}
% ======================
%#######################
\subsection{Incompleteness Bias}
\label{sec:incompleteness_bias}
The \simstack\ technique returns an unbiased estimate of the flux density when \emph{all} potentially correlated sources are accounted for.  
When sources are missing, either because they are too dust-obscured or intrinsically faint to detect, their unaccounted-for flux may potentially bias the result.  
While the UDS is an exceptionally deep survey, with greater than 90\% completeness at $z \sim 3.5$ and log($M/ \rm M_{\odot})=10.5$, it is possible that the absence of lower-mass galaxies, particularly at higher redshifts, could affect the results.  
We explore the severity of a potential bias with the following simulation. 

Restricting ourselves to star-forming galaxies, 
we first simultaneously stack all mass bins at a single redshift and record their fluxes.    
Next, we remove the lowest mass bin and stack again, 
comparing the fluxes in the remaining bins to those of the initial stacked fluxes.  
We repeat this until only the highest mass bin remains.     
Results are as we might expect:   
the removal of bins increases the stacked fluxes of the remaining bins in a systematic way, from negligibly little for one missing bin, to as much as 3-22\% for the highest-mass bin alone; and that the level of the bias increases with beam size.   
For the case of two mass bins removed---roughly equivalent to the scenario of our highest redshift bin---the bias ranges over 0.2--4.0\%, from smallest to largest beam size.  

What does this mean for resolving the background?  
Although missing sources boost the fluxes of those remaining, 
the sum of their missing flux counteracts the boosting, resulting in a marginal bias.  
Our simulations confirm this, with the removal of up to three bins having negligible impact. 

%#######################
\subsection{Contribution from Quiescent Galaxies}
\label{sec:quiescent}
% ######################

Here we briefly discuss the small contribution ($\sim 5\%$) to the
total resolved CIB from galaxies classified as quiescent.  The
majority of this contribution originates at $z \gsim 2$, in contrast
with the rest of the CIB, which comes mostly from star-forming galaxies
at $z=0$--2.  Indeed, at $z \gsim 2$ the most massive 
quiescent galaxies ($M = 10^{11-12}\,\rm M_{\odot}$) are ULIRGs, and
the next most massive ($M = 10^{10-11}\,\rm M_{\odot}$) are LIRGs.

The simplest  explanation is that star-forming
galaxies are being misclassified as quiescent galaxies,  particularly 
 at $z \gsim 2.5$, where their signal-to-noise in the optical is lower, and their 
colors thus more likely to be scattered into the quiescent part of the \emph{UVJ} plane.  
But dig a little deeper and that scenario alone seems unlikely, as to achieve ULIRG-like luminosities \emph{on average}, either \emph{all} quiescent galaxies at the highest redshifts must be misclassified ULIRGS---i.e., \emph{no} passive quiescent galaxies past redshift of $\sim 2.5$---or that the fraction that is misclassified would need to be incredibly luminous and abundant (e.g., $L\sim 2$--$4\times 10^{12}\, \rm L_{\odot}$ for 50\% contamination, $L\sim 4$--$8\times 10^{12}\, \rm L_{\odot}$ for 25\% contamination, etc.);  
 both of which would be wildly inconsistent with the latest measurements of the stellar mass function \citep{muzzin2013, ilbert2013}.  

Alternatively, it might be that interlopers misclassified in redshift as well as color are boosting the flux.  
This would result in a slightly hotter SED in the rest frame---which we do see---and significant boosting of $\nu I_{\nu}$ without the need for excessively luminous sources.   
However, we have already seen that as a result of negative $K$-correction, flux densities are roughly flat with redshift  in the submillimeter bands (e.g., Figure~\ref{fig:stacked_fluxes}), so that only relatively local (i.e., $z\lsim 0.5$) sources would be capable of boosting the flux densities enough to account for the increased luminosities.  But as that is also where photo-$z$ are most reliable \citep[e.g.,][]{brammer2008, quadri2010}, it unlikely that this is taking place.     

Aside from measurement errors, it is worth noting that there are
possible physical explanations for the FIR emission.  
Because the star formation in quiescent galaxies can only have shut off very recently
at these redshifts, there may still be a strong interstellar radiation
field heating the dust.  
For example, \citet{fumagalli2013} explore whether not supposedly quiescent galaxies (also selected with the \emph{UVJ} technique) are indeed dead.  They split up their sample into five redshift bins and select only quiescent galaxies with stellar masses greater than log($M/\rm M_{\odot}) > 10.3$, finding log($L/L_{\odot}) \sim 9.5$--10.5 over the redshift range $0.3 < z < 2.5$.   These luminosities are largely consistent with our findings---both in amplitude and in the evolution of that amplitude---but fall short of the redshift range where we find passive galaxies with ULIRG-like luminosities.  
They conclude that circumstellar dust from AGB stars \citep[e.g.,][]{knapp1992,lancon2002,piovan2003}, and cirrus heated by old stellar populations \citep[e.g., the so-called cirrus heating][]{kennicutt1998, salim2009}, are probably responsible for much of the IR emission in their sample, and that the star-formation rates in these galaxies is some 20--40 times lower than in their star-forming counterparts.  

Another possibility is that there is FIR emission associated with embedded AGN in highly obscured red galaxies at high redshift \citep[e.g.,][]{polletta2006,polletta2008, daddi2007b,alexander2008,kirkpatrick2012}.   AGN are found to have luminosities log($L/\rm L_{\odot})> 12$--13 at $z>1$ \citep[e.g.,][]{pope2008,dai2012}, which would 
be enough to explain the rising luminosity with redshift, particularly since the evolution of the AGN 
fraction follows a similar trend \citep[e.g.,][]{alexander2005}.    
We explore this scenario in more detail in the next section.  

Finally, it might just be possible that increasing levels of star formation is present but so heavily obscured that 
 the contribution from young stars to the rest-frame \emph{J}-band luminosity is small.  
 There are examples of  high-redshift passive galaxies with 24\rmicron\ flux 
suggesting excess FIR emission, \citep[e.g.,][]{marchesini2010,perez2008};   
 as well as optically selected red-and-dead galaxies  stacked in the submillimeter 
showing excess flux, possibly indicating star-formation \citep{viero2012}.  
%#######################
\subsection{Active Galactic Nuclei and the High $\chi^2$ Sample}
\label{sec:agn}
%#######################
AGN have been found to contribute to the infrared luminosities of both passive and star-forming galaxies  at levels between 3 and 37\% \citep[e.g.,][]{pope2008,mullaney2011,juneau2013}.  While submillimeter colors alone are inadequate for identifying AGN \citep[e.g.,][]{hatziminaoglou2010,kirkpatrick2013},  
the mid-infrared has been established as a sensitive probe of galaxy type \citep[e.g.,][]{brandl2006,farrah2007}, 
with the presence of an exponentially rising SED a robust signature of AGN \citep[][]{lacy2004,richards2006,hickox2007,assef2010,mullaney2011,stern2012,feltre2013}.  
When spectra are not available, it has been shown that higher 22--24\rmicron\ flux densities correlate with higher AGN fractions \citep[][]{daddi2007b,lee2010,murphy2011,bridge2013}.    

We look for evidence of AGN in the roughly 2{,}700 sources that were otherwise poorly fit with standard SEDs by adding an AGN template \citep[e.g.,][]{assef2010} to the fit, 
and find that approximately half of the fits are significantly improved.   
Next, we stack them separately from---but simultaneously with---the star-forming and quiescent galaxies to look for further indications of AGN.  
 We find that the contribution to the CIB from these sources is $\lsim 4\%$, and is distributed over the same redshift range as the star-forming galaxies.  
Additionally, we find that the SEDs are on average $\sim 10\, \rm K$ hotter, an indication that emission from hot dust around the AGN \citep[e.g.,][]{rowan1989,lacy2004} is contributing to the thermal SED at shorter wavelengths \citealp[(e.g.,][]{kirkpatrick2012}\citealp[; although also see][]{feltre2013}). 
We also find that the 24\rmicron\ flux density exceeds that of the other galaxies of similar mass, again, a strong signature of AGN \citep[e.g.,][]{daddi2007b,lee2010}.

While intriguing, fully understanding the contribution from AGN is complicated and subtle, and lies beyond the scope of this paper.  
It is worth noting that because AGN emission should exist at varying levels in all galaxies \citep[e.g.,][]{juneau2013}, the total contribution to the CIB attributable to AGN may in fact be significant.   
A full treatment of this class of sources requires exploring this threshold with better mid-infrared diagnostics and fits using a full set of AGN templates will be explored in detail in Moncelsi et al.\@ (in prep.).

%#######################
\subsection{Where is the Missing CIB?}
\label{sec:missing}
%#######################
Based on our best-fit SEDs, we resolve 
$2.0\pm0.5$, $5.2\pm1.2$, $7.4\pm1.7$, $10.7\pm2.5$, $8.4\pm1.8$, $4.54\pm1.0$, $1.7\pm 0.3$, and $0.1,\pm0.02\,\rm nW\,m^{-2}\,sr^{-1}$ at 
24, 70, 100, 160, 250, 350, 500, and 1100\rmicron, respectively.   
This accounts for between  ($45 \pm 8$)\% and ($80 \pm 17$)\% of the CIB, with the lowest fractions resolved at the  longest (AzTEC) wavelengths, and the highest fractions in the three SPIRE bands.    
This is similar to the resolved fraction found by \citet{marsden2009}---although they likely suffered from a bias due to neglecting clustering within the large BLAST beams \citep{pascale2008}---and by Vieira et al.\@ (in prep.)\@ and \citet{bethermin2012b} at far-infrared/submillimeter wavelengths. 

Between $(55\pm 8)$\% and $(20\pm 17)$\% of the CIB thus appears to be unaccounted for by our \emph{K}-selected sample.  
It is notable that the measurements for the absolute CIB are themselves uncertain at the $\sim 25\%$ level, so that formally our measurements could be said to be consistent with the CIB in any one band.  However, considering that they fall below the nominal CIB in every band, the quoted percentage should be a fair estimate.    

Besides the possibility that the absolute CIB is not entirely of extragalactic origin, the two most obvious candidates for the missing fraction are: 1) a large number of low-mass, intrinsically faint sources undetected by the UDS survey; and 2) IR luminous, potentially massive but unusually dust-obscured sources also missed by the UDS selection.  We will now discuss these possibilities.  

%#######################
\subsubsection{Low-Mass Faint Sources}
\label{sec:faint_sources}
%#######################
Our sample is  selected at $K_{\rm AB} \leq 24$, reaching a source density on the sky of $\sim 36\, \rm arcmin^{-2}$.  
Fainter, lower-mass sources missed by the catalog certainly exist \citep[i.e., the faint end of the luminosity function, e.g.,][]{muzzin2013,ilbert2013}, but could they contribute enough intensity to make up the remaining CIB?  

If we assume  that there are as many undetected sources  as there are detected, 
 and that at, say, 500\rmicron, their flux densities are 0.1\,mJy (i.e., just below that of the current lowest mass bin) at all redshifts (not unreasonable given negative $K$-correction, e.g., see Figure~\ref{fig:stacked_fluxes}), that would add $\sim 0.25\pm 0.05\, \rm nW\, m^{-2}\, sr^{-1}$, i.e., another $\sim (10\pm 2)\%$, to the CIB.  
This behavior mirrors the behavior demonstrated in  \S~\ref{sec:resolved_cib}, where aggressively correcting for completeness brought the CIB to within 5\% of the nominal total.  
We thus conclude that some fraction is probably due to the faint population, though without a full simulation, or deeper data, it is difficult to say how much.    

%\newpage
%#######################
\subsubsection{Dust-Obscured Sources}
\label{sec:missing_sources}
%#######################
There exists sources---for example submillimeter galaxies---that are incredibly luminous in the infrared (e.g., $L_{\rm IR} > 10^{12}\, \rm L_{\odot} $) but so dust-obscured that they are often very faint at UV/optical wavelengths \citep[][]{smail1997,barger1998,hughes1998,blain2002}.    
For example, \citet[][]{dey1999}, observing a log($L/\rm L_{\odot})=12.8$ ULIRG at $z=1.44$ with the Wide Field Planetary Camera 2 \citep[WFPC2;][]{holtzman1995} on the \emph{Hubble Space Telescope} with the F814W filter ($\lambda=7930\, \rm \AA$) measured a magnitude of $24.6\pm 0.1$\,[AB]; i.e., this source would probably not have made our cut.    
These sources tend be irregular galaxies---likely due to a recent or ongoing merger \citep[][]{hernquist1989,hayward2013}---forming stars in bursts rather than at a steady rate governed by the infrared \lq\lq main sequence\rq\rq\ for star-forming galaxies \citep[e.g.,][]{noeske2007,elbaz2011}.   

A hint that luminous but obscured sources might be going undetected comes from Figure~\ref{fig:lcib}, where the resolved fraction of ULIRGs is in disagreement with the \citet{bethermin2011} model, while less-luminous sources appear to be in quite good agreement.  Since the sum of the three colored curves resolves the full CIB, it may well be that this missing component is rather significant.

Likely better tracers of highly dust-obscured star-forming galaxies are 24\rmicron\ sources, which, as already mentioned, have been used by several groups as positional priors.  
For example, \citet{dole2006} found that 24\rmicron\ selected sources make up 70\% of the CIB at 70 and 100\rmicron;   
while \citet{berta2011} found they could explain between 58 and 64\% of the CIB at 100 and 160\rmicron; 
and \citet{bethermin2012b} resolve 70\% of the background at 250, 350, and 500\rmicron.  
However, a considerable fraction of the emission at 24\rmicron\ is shown to originate from galaxies hosting AGN (e.g., Vieira et al.\@ in prep.).  

As we discussed in \S~\ref{sec:agn}, the inclusion of high $\chi^2$ sources increased  the stacked 24\rmicron\ flux density more than at other bands.  
And although we resolve about the same amount of the CIB, it does not mean we are resolving the same sources.   It is possible that other faint AGN are being missed, particularly at high redshift, where the AGN contribution dominates.  
The potential thus exists that we are missing a significant contribution from dust-obscured galaxies.  
In the future, identifying 24\rmicron\ and other sources missed by our \emph{K}-selected catalog---and stacking them simultaneously with our optically selected sources---would be a natural extension of this work.

% ======================
\section{Summary}
\label{sec:conclusion}
% ======================
A complete understanding of galaxies and galaxy evolution requires characterizing their full SEDs, a goal which has to-date been technically challenging, as galaxies in the submillimeter are mostly unresolved.  And for those that are resolved, large instrumental beams makes counterpart identification laborious.    
Here we attempted to statistically connect galaxies selected in the \emph{K}-band to those that make up the CIB by stacking sources grouped into bins of mass, redshift, and color---presenting and making public our own \simstack\ algorithm.    

We found that between ($45 \pm 8$)\% and ($80 \pm 17$)\% of the total CIB is resolved by our catalog, and that the bulk originates from $z=0$--2 and log($M/\rm M_{\odot}) = 10.0$--11.0.  
Sources at higher redshifts contribute more at longer wavelengths, a consequence of the peak of the SED redshifting into the submillimeter/millimeter bands (i.e., negative $K$-correction).  
Higher-mass LIRGs and ULIRGs are seen to contribute at longer wavelengths (i.e., higher redshift), which shifts to lower mass LIRGs and \lq\lq normal\rq\rq\ star-forming galaxies at shorter wavelengths. 

%SOMETHING ABOUT LUMINOSITIES AND DENSITIES.
We find that the luminosities of all galaxies rises rapidly with increasing redshift, but whereas they continue to rise until $z=4$ for the most massive galaxies, there is an apparent turnover at $z\gsim 2$ in galaxies of log($M/\rm M_{\odot}) \lsim 10.5$.   
We further find that while galaxies identified as quiescent by their colors have very little emission at low redshift,  beyond $z\ge 2$ they evolve even more rapidly than their star-forming counterparts.  
%Similarly, the luminosity densities of the most massive star-forming and quiescent galaxies rises rapidly
This is likely due, to some extent, to limitations in the \emph{UVJ} color selection at higher redshifts resulting from noisy photometry, though the origin of the entirety of this flux is still not certain, and will be the subject of future work.    
Lastly, galaxies whose photometry is poorly fit by standard SEDs appear to have a clear association with AGN;  
having comparable luminosities but higher temperatures as star-forming galaxies of the same stellar mass.    

%SOMETHING ABOUT TEMPERATURES
%HOW ABOUT THE FITS?  

Our work pushes the boundaries of characterizing low mass and high-redshift optical galaxies in the far-infrared/submillimeter, but nevertheless, more can be done.  Though the UDS is quite deep, deeper and/or wider fields and new catalogs are now coming online \citep[e.g.,][]{muzzin2013b,ilbert2013,viero2013c}.  That, combined with future ancillary data over large areas and to longer wavelengths, and with deeper near-infrared imaging and 24\rmicron\ catalogs, should allow us to reach the coveted goal of resolving the full cosmic background.  

\begin{acknowledgments}
\section*{Acknowledgments}
The authors warmly thank  Duncan Hanson, Phil Korngut, Zak Staniszewski, and Yoshihiro Ueda.  
We also thank the anonymous referee, whose careful comments greatly improved this paper.  
Much credit belongs to C. Barth Netterfield and Enzo Pascale for inspiring the simultaneous stacking algorithm, and to whom we are thankful.  
\input{spire_acknowledgement.tex}
RJA was supported by Gemini-CONICYT grant number 32120009.
\end{acknowledgments}

%\newpage

\bibliographystyle{apj}
\bibliography{refs}

%APPENDIX
\appendix

\section{A. Tables}
\label{sec:tables}
\input{t1.tex}
\input{t2.tex}
\input{t3.tex}
\input{t4.tex}
\input{t5.tex}

% ======================
% ======================

\end{document}

%% file: table1.tex
\begin{table}
 \centering
  \caption{Nominal and Effective Beam Properties.}
     \begin{tabular}{l||ccc}
       % \hline
       %Wavelength  & Nominal FWHM & Nominal Solid Angle & Effective FWHM & Effective Solid Angle \\
       Band  & $\rm FWHM_{nom}$ & $\rm FWHM_{eff}$ & $\rm Area_{\rm eff}$   \\
       ($\micro$m) &   (arcsec)  & (arcsec)  & (steradians)   \\
       \hline
       24 &  6.0 &  6.3 &     $1.548\times 10^{-9}$  \\
       70 & 18.0 &    19.3 & $1.296\times 10^{-8}$  \\
       100 & 7.4 & 7.0 & $1.305\times 10^{-9}$   \\
       160 & 11.3 & 11.2 & $3.341\times 10^{-9}$   \\
       250 & 18.1 & 17.6 & $0.994\times 10^{-8}$   \\
       350 & 25.2 & 23.9 & $1.765\times 10^{-8}$   \\
       500 & 36.6 & 35.2 & $3.730\times 10^{-8}$    \\
       1100 & 30.0 & 30.0 & $3.179\times 10^{-8}$    \\
       %   \hline
         \end{tabular}
	 %   \caption{Nominal and Effective Beam Properties.}
	  \label{tab:beams}
	  \end{table}

%% file: table2.tex
\begin{table}
  \centering
    \caption{Pearson correlation matrix for all bands under analysis.}
  \begin{tabular}{c|cccccccc}
  % \hline
  Band  &   &      &      &      &      &   &     & \\
 ($\micro$m) & 24   & 70   & 100 & 160 & 250 & 350 & 500 & 1100\\
   \hline
24    &1 &       0.23 &0.33  & 0.32 & 0.28 & 0.17 &  0.07  &  0.10 \\
70    &   &    1    & 0.19 & 0.24 &  0.23 & 0.14  & 0.06 & 0.08 \\
100 &    &          &  1     &    0.28 & 0.21 & 0.11 & 0.04 & 0.05 \\
160  &   &          &          &    1      &  0.35 & 0.23 & 0.10  & 0.13 \\
250  &   &          &          &           &   1    & 0.37 &  0.18  &  0.28 \\
350  &   &          &          &          &          &    1    &  0.20  & 0.33  \\
500  &   &          &          &         &           &            & 1       & 0.23  \\
1100 &  &          &         &          &           &           &           &  1      \\
  % \hline
  \end{tabular}
  \label{tab:covariance_matrix}
\end{table}

%% file: table3.tex
\begin{table*}
 \scriptsize 
 \centering
\caption{Total Stacked Intensities}
 \begin{tabular}{cccccc}
 \hline
 \hline
$ $ & \multicolumn{3}{c}{ This Work } &  \multicolumn{2}{c}{ Absolute Measurements } \\
{\bf Band} & {\bf Total Stacking} & {\bf Completeness Corrected } & {\bf Total Model SEDs} & {\bf Absolute CIB } & {\bf Reference}   \\ 
 ($\micro \rm m$) & ($\rm nW\, m^{-2}\, sr^{-1}$) & ($\rm nW\, m^{-2}\, sr^{-1}$) & ($\rm nW\, m^{-2}\, sr^{-1}$) & ($\rm nW\, m^{-2}\, sr^{-1}$) &  \\ 
 \hline
24 & $1.84 \pm 0.05$ $(64 \pm 1\%)$ & $1.87 \pm 0.05$ $(65 \pm 1\%)$  & $ 1.99 \pm 0.46$ $(69 \pm 15\%)$ & $2.86 \pm 0.17 $ & \citet{bethermin2010} \\
70 & $3.31 \pm 0.20$ $(50 \pm 3\%)$ & $3.33 \pm 0.20$ $(50 \pm 3\%)$  & $ 5.17 \pm 1.20$ $(78 \pm 18\%)$ & $6.60 \pm 0.70 $ & \citet{bethermin2010} \\
100 & $8.74 \pm 0.38$ $(69 \pm 3\%)$ & $8.84 \pm 0.39$ $(70 \pm 3\%)$  & $ 7.35 \pm 1.72$ $(58 \pm 13\%)$ & $12.60 \pm 4.00 $ & \citet{berta2011} \\
160 & $9.43 \pm 0.63$ $(69 \pm 4\%)$ & $9.57 \pm 0.64$ $(70 \pm 4\%)$  & $ 10.66 \pm 2.50$ $(78 \pm 18\%)$ & $13.60 \pm 2.50 $ & \citet{berta2011} \\
250 & $7.00 \pm 0.34$ $(67 \pm 3\%)$ & $7.14 \pm 0.35$ $(68 \pm 3\%)$  & $ 8.39 \pm 1.83$ $(80 \pm 17\%)$ & $10.40 \pm 2.30 $ & \citet{lagache2000} \\
350 & $4.38 \pm 0.22$ $(67 \pm 3\%)$ & $4.50 \pm 0.23$ $(69 \pm 3\%)$  & $ 4.54 \pm 0.92$ $(69 \pm 14\%)$ & $6.50 \pm 1.60 $ & \citet{lagache2000} \\
500 & $1.84 \pm 0.10$ $(70 \pm 3\%)$ & $1.91 \pm 0.11$ $(73 \pm 4\%)$  & $ 1.69 \pm 0.32$ $(65 \pm 12\%)$ & $2.60 \pm 0.60 $ & \citet{lagache2000} \\
1100 & $0.06 \pm 0.01$ $(34 \pm 2\%)$ & $0.07 \pm 0.01$ $(35 \pm 2\%)$  & $ 0.09 \pm 0.02$ $(45 \pm 8\%)$ & $0.19 \pm 0.04 $ & \citet{lagache2000} \\
\hline
\end{tabular}
\tablecomments{Errors estimated with a extended bootstrap technique described in \S~\ref{sec:errors}. In parentheses are the percentages of the total CIB resolved, as measured at 24--160\rmicron\ by MIPS; and at 250--1100\rmicron\ by FIRAS.}
\label{tab:cib_totals}
\vspace{2mm}
\end{table*}

%% file: spire_acknowledgement.tex
SPIRE has been developed by a consortium of institutes led
by Cardiff Univ. (UK) and including: Univ. Lethbridge (Canada);
NAOC (China); CEA, LAM (France); IFSI, Univ. Padua (Italy);
IAC (Spain); Stockholm Observatory (Sweden); Imperial College
London, RAL, UCL-MSSL, UKATC, Univ. Sussex (UK); and Caltech,
JPL, NHSC, Univ. Colorado (USA). This development has been
supported by national funding agencies: CSA (Canada); NAOC
(China); CEA, CNES, CNRS (France); ASI (Italy); MCINN (Spain);
SNSB (Sweden); STFC, UKSA (UK); and NASA (USA).

%% file: t1.tex
\begin{table*}
\tiny
 \centering
\caption{Stacked Flux Densities of Star-Forming Galaxies}
 \begin{tabular}{cccccc}
 \hline
$\lambda $ & \multicolumn{5}{c}{${\bf  z = 0.0 }$--${\bf 0.5 }$}  \\
($\micro \rm m$) &  $  {\rm log}(M/{\rm M}_{\odot}) = 9.0$--$10.0 $  &  $  {\rm log}(M/{\rm M}_{\odot}) = 10.0$--$11.0 $  &  $  {\rm log}(M/{\rm M}_{\odot}) = 11.0$--$12.0 $  &  $  {\rm log}(M/{\rm M}_{\odot}) = 9.0$--$9.5 $  &  $  {\rm log}(M/{\rm M}_{\odot}) = 9.5$--$10.0 $  \\
 \hline
$ 24$ & $ (6.7 \pm 0.7)\times 10^{-2} $  &  $ (2.1 \pm 0.2)\times 10^{-1} $  &  $ (4.8 \pm 0.3)\times 10^{-1} $  &  $ (3.9 \pm 0.5)\times 10^{-1} $  &  $ (6.8 \pm 0.0)\times 10^{-1} $  \\ 
$ 70$ & $ (7.0 \pm 1.3)\times 10^{-1} $  &  $ (2.4 \pm 0.2)\times 10^{0} $  &  $ (4.8 \pm 0.5)\times 10^{0} $  &  $ (3.8 \pm 0.8)\times 10^{0} $  &  $ (1.4 \pm 28.0)\times 10^{1} $  \\ 
$ 100$ & $ (1.8 \pm 0.1)\times 10^{0} $  &  $ (5.8 \pm 0.4)\times 10^{0} $  &  $ (1.3 \pm 0.1)\times 10^{1} $  &  $ (1.1 \pm 0.1)\times 10^{1} $  &  $ (2.6 \pm 40.0)\times 10^{1} $  \\ 
$ 160$ & $ (2.4 \pm 0.7)\times 10^{0} $  &  $ (7.2 \pm 1.0)\times 10^{0} $  &  $ (1.7 \pm 0.2)\times 10^{1} $  &  $ (1.8 \pm 0.3)\times 10^{1} $  &  $ (3.9 \pm 36.0)\times 10^{1} $  \\ 
$ 250$ & $ (1.8 \pm 0.2)\times 10^{0} $  &  $ (5.5 \pm 0.4)\times 10^{0} $  &  $ (1.3 \pm 0.1)\times 10^{1} $  &  $ (1.5 \pm 0.1)\times 10^{1} $  &  $ (3.5 \pm 0.2)\times 10^{1} $  \\ 
$ 350$ & $ (1.1 \pm 0.2)\times 10^{0} $  &  $ (3.1 \pm 0.3)\times 10^{0} $  &  $ (7.3 \pm 0.6)\times 10^{0} $  &  $ (9.4 \pm 1.1)\times 10^{0} $  &  $ (1.5 \pm 0.2)\times 10^{1} $  \\ 
$ 500$ & $ (4.6 \pm 1.9)\times 10^{-1} $  &  $ (1.6 \pm 0.3)\times 10^{0} $  &  $ (3.6 \pm 0.5)\times 10^{0} $  &  $ (4.3 \pm 1.0)\times 10^{0} $  &  $ (2.1 \pm 1.7)\times 10^{0} $  \\ 
$ 1100$ & ---  &  $ (9.2 \pm 4.9)\times 10^{-2} $  &  $ (1.6 \pm 0.7)\times 10^{-1} $  &  $ (4.3 \pm 1.5)\times 10^{-1} $  &  $ (3.2 \pm 0.7)\times 10^{-1} $  \\ 
 \hline
$\lambda $ & \multicolumn{5}{c}{${\bf  z = 0.5 }$--${\bf 1.0 }$}  \\
($\micro \rm m$) &  $  {\rm log}(M/{\rm M}_{\odot}) = 9.0$--$10.0 $  &  $  {\rm log}(M/{\rm M}_{\odot}) = 10.0$--$11.0 $  &  $  {\rm log}(M/{\rm M}_{\odot}) = 11.0$--$12.0 $  &  $  {\rm log}(M/{\rm M}_{\odot}) = 9.0$--$9.5 $  &  $  {\rm log}(M/{\rm M}_{\odot}) = 9.5$--$10.0 $  \\
 \hline
$ 24$ & $ (1.4 \pm 0.1)\times 10^{-2} $  &  $ (8.5 \pm 0.4)\times 10^{-2} $  &  $ (2.0 \pm 0.1)\times 10^{-1} $  &  $ (3.6 \pm 0.1)\times 10^{-1} $  &  $ (2.1 \pm 0.3)\times 10^{-1} $  \\ 
$ 70$ & ---  &  $ (6.8 \pm 0.7)\times 10^{-1} $  &  $ (1.1 \pm 0.1)\times 10^{0} $  &  $ (2.0 \pm 0.2)\times 10^{0} $  &  $ (1.0 \pm 0.5)\times 10^{0} $  \\ 
$ 100$ & $ (3.7 \pm 0.4)\times 10^{-1} $  &  $ (1.8 \pm 0.1)\times 10^{0} $  &  $ (4.0 \pm 0.1)\times 10^{0} $  &  $ (6.1 \pm 0.3)\times 10^{0} $  &  $ (2.3 \pm 0.8)\times 10^{0} $  \\ 
$ 160$ & $ (4.8 \pm 3.0)\times 10^{-1} $  &  $ (2.7 \pm 0.4)\times 10^{0} $  &  $ (6.4 \pm 0.7)\times 10^{0} $  &  $ (1.0 \pm 0.1)\times 10^{1} $  &  $ (7.6 \pm 3.0)\times 10^{0} $  \\ 
$ 250$ & $ (5.3 \pm 0.8)\times 10^{-1} $  &  $ (2.7 \pm 0.1)\times 10^{0} $  &  $ (6.7 \pm 0.3)\times 10^{0} $  &  $ (1.1 \pm 0.1)\times 10^{1} $  &  $ (8.9 \pm 1.4)\times 10^{0} $  \\ 
$ 350$ & $ (4.1 \pm 0.9)\times 10^{-1} $  &  $ (1.8 \pm 0.1)\times 10^{0} $  &  $ (4.8 \pm 0.3)\times 10^{0} $  &  $ (8.6 \pm 0.5)\times 10^{0} $  &  $ (7.2 \pm 1.2)\times 10^{0} $  \\ 
$ 500$ & $ (2.5 \pm 0.8)\times 10^{-1} $  &  $ (6.7 \pm 1.2)\times 10^{-1} $  &  $ (2.1 \pm 0.2)\times 10^{0} $  &  $ (4.8 \pm 0.4)\times 10^{0} $  &  $ (3.7 \pm 0.9)\times 10^{0} $  \\ 
$ 1100$ & ---  &  $ (1.3 \pm 2.0)\times 10^{-2} $  &  $ (1.7 \pm 0.3)\times 10^{-1} $  &  $ (2.9 \pm 0.7)\times 10^{-1} $  &  $ (5.2 \pm 1.7)\times 10^{-1} $  \\ 
 \hline
$\lambda $ & \multicolumn{5}{c}{${\bf  z = 1.0 }$--${\bf 1.5 }$}  \\
($\micro \rm m$) &  $  {\rm log}(M/{\rm M}_{\odot}) = 9.0$--$10.0 $  &  $  {\rm log}(M/{\rm M}_{\odot}) = 10.0$--$11.0 $  &  $  {\rm log}(M/{\rm M}_{\odot}) = 11.0$--$12.0 $  &  $  {\rm log}(M/{\rm M}_{\odot}) = 9.0$--$9.5 $  &  $  {\rm log}(M/{\rm M}_{\odot}) = 9.5$--$10.0 $  \\
 \hline
$ 24$ & ---  &  $ (2.7 \pm 0.2)\times 10^{-2} $  &  $ (9.2 \pm 0.5)\times 10^{-2} $  &  $ (1.6 \pm 0.1)\times 10^{-1} $  &  $ (2.7 \pm 0.2)\times 10^{-1} $  \\ 
$ 70$ & ---  &  $ (4.7 \pm 5.3)\times 10^{-2} $  &  $ (4.9 \pm 0.8)\times 10^{-1} $  &  $ (1.1 \pm 0.1)\times 10^{0} $  &  $ (1.5 \pm 0.3)\times 10^{0} $  \\ 
$ 100$ & $ (6.5 \pm 3.6)\times 10^{-2} $  &  $ (6.3 \pm 0.5)\times 10^{-1} $  &  $ (1.9 \pm 0.1)\times 10^{0} $  &  $ (3.5 \pm 0.1)\times 10^{0} $  &  $ (4.0 \pm 0.4)\times 10^{0} $  \\ 
$ 160$ & ---  &  $ (8.6 \pm 3.5)\times 10^{-1} $  &  $ (3.5 \pm 0.5)\times 10^{0} $  &  $ (6.8 \pm 0.8)\times 10^{0} $  &  $ (1.0 \pm 0.2)\times 10^{1} $  \\ 
$ 250$ & $ (5.5 \pm 8.1)\times 10^{-2} $  &  $ (1.4 \pm 0.1)\times 10^{0} $  &  $ (4.2 \pm 0.2)\times 10^{0} $  &  $ (8.8 \pm 0.4)\times 10^{0} $  &  $ (1.4 \pm 0.1)\times 10^{1} $  \\ 
$ 350$ & ---  &  $ (1.3 \pm 0.1)\times 10^{0} $  &  $ (3.7 \pm 0.2)\times 10^{0} $  &  $ (8.0 \pm 0.4)\times 10^{0} $  &  $ (1.2 \pm 0.1)\times 10^{1} $  \\ 
$ 500$ & ---  &  $ (5.8 \pm 1.0)\times 10^{-1} $  &  $ (2.0 \pm 0.2)\times 10^{0} $  &  $ (4.5 \pm 0.3)\times 10^{0} $  &  $ (7.4 \pm 0.8)\times 10^{0} $  \\ 
$ 1100$ & ---  &  $ (2.2 \pm 1.8)\times 10^{-2} $  &  $ (1.3 \pm 0.3)\times 10^{-1} $  &  $ (4.4 \pm 0.5)\times 10^{-1} $  &  $ (3.7 \pm 1.0)\times 10^{-1} $  \\ 
 \hline
$\lambda $ & \multicolumn{5}{c}{${\bf  z = 1.5 }$--${\bf 2.0 }$}  \\
($\micro \rm m$) &  $  {\rm log}(M/{\rm M}_{\odot}) = 9.0$--$10.0 $  &  $  {\rm log}(M/{\rm M}_{\odot}) = 10.0$--$11.0 $  &  $  {\rm log}(M/{\rm M}_{\odot}) = 11.0$--$12.0 $  &  $  {\rm log}(M/{\rm M}_{\odot}) = 9.0$--$9.5 $  &  $  {\rm log}(M/{\rm M}_{\odot}) = 9.5$--$10.0 $  \\
 \hline
$ 24$ & ---  &  $ (1.7 \pm 0.1)\times 10^{-2} $  &  $ (8.7 \pm 0.4)\times 10^{-2} $  &  $ (1.7 \pm 0.1)\times 10^{-1} $  &  $ (3.0 \pm 0.3)\times 10^{-1} $  \\ 
$ 70$ & ---  &  ---  &  $ (2.7 \pm 0.9)\times 10^{-1} $  &  $ (3.9 \pm 1.3)\times 10^{-1} $  &  $ (1.6 \pm 0.2)\times 10^{0} $  \\ 
$ 100$ & $ (1.4 \pm 0.4)\times 10^{-1} $  &  $ (2.4 \pm 0.6)\times 10^{-1} $  &  $ (1.3 \pm 0.1)\times 10^{0} $  &  $ (2.1 \pm 0.1)\times 10^{0} $  &  $ (3.8 \pm 0.3)\times 10^{0} $  \\ 
$ 160$ & ---  &  $ (5.0 \pm 4.4)\times 10^{-1} $  &  $ (2.3 \pm 0.6)\times 10^{0} $  &  $ (4.1 \pm 0.9)\times 10^{0} $  &  $ (7.5 \pm 1.8)\times 10^{0} $  \\ 
$ 250$ & ---  &  $ (8.6 \pm 1.3)\times 10^{-1} $  &  $ (3.1 \pm 0.2)\times 10^{0} $  &  $ (6.5 \pm 0.4)\times 10^{0} $  &  $ (1.3 \pm 0.1)\times 10^{1} $  \\ 
$ 350$ & ---  &  $ (8.6 \pm 1.4)\times 10^{-1} $  &  $ (3.0 \pm 0.2)\times 10^{0} $  &  $ (6.4 \pm 0.4)\times 10^{0} $  &  $ (1.3 \pm 0.1)\times 10^{1} $  \\ 
$ 500$ & ---  &  $ (6.0 \pm 1.3)\times 10^{-1} $  &  $ (1.9 \pm 0.2)\times 10^{0} $  &  $ (4.5 \pm 0.3)\times 10^{0} $  &  $ (8.9 \pm 0.7)\times 10^{0} $  \\ 
$ 1100$ & ---  &  $ (1.4 \pm 2.2)\times 10^{-2} $  &  $ (1.4 \pm 0.4)\times 10^{-1} $  &  $ (3.4 \pm 0.6)\times 10^{-1} $  &  $ (9.6 \pm 1.1)\times 10^{-1} $  \\ 
 \hline
$\lambda $ & \multicolumn{5}{c}{${\bf  z = 2.0 }$--${\bf 2.5 }$}  \\
($\micro \rm m$) &  $  {\rm log}(M/{\rm M}_{\odot}) = 9.0$--$10.0 $  &  $  {\rm log}(M/{\rm M}_{\odot}) = 10.0$--$11.0 $  &  $  {\rm log}(M/{\rm M}_{\odot}) = 11.0$--$12.0 $  &  $  {\rm log}(M/{\rm M}_{\odot}) = 9.0$--$9.5 $  &  $  {\rm log}(M/{\rm M}_{\odot}) = 9.5$--$10.0 $  \\
 \hline
$ 24$ & ---  &  $ (1.1 \pm 1.6)\times 10^{-3} $  &  $ (5.2 \pm 0.4)\times 10^{-2} $  &  $ (1.3 \pm 0.1)\times 10^{-1} $  &  $ (2.7 \pm 0.1)\times 10^{-1} $  \\ 
$ 70$ & ---  &  ---  &  $ (9.4 \pm 9.6)\times 10^{-2} $  &  $ (4.1 \pm 1.3)\times 10^{-1} $  &  $ (1.3 \pm 0.2)\times 10^{0} $  \\ 
$ 100$ & $ (2.1 \pm 0.8)\times 10^{-1} $  &  ---  &  $ (6.0 \pm 0.9)\times 10^{-1} $  &  $ (1.3 \pm 0.1)\times 10^{0} $  &  $ (2.7 \pm 0.2)\times 10^{0} $  \\ 
$ 160$ & ---  &  ---  &  $ (1.3 \pm 0.7)\times 10^{0} $  &  $ (3.5 \pm 1.0)\times 10^{0} $  &  $ (6.8 \pm 1.4)\times 10^{0} $  \\ 
$ 250$ & ---  &  $ (7.5 \pm 15.0)\times 10^{-2} $  &  $ (2.0 \pm 0.2)\times 10^{0} $  &  $ (5.2 \pm 0.3)\times 10^{0} $  &  $ (1.2 \pm 0.1)\times 10^{1} $  \\ 
$ 350$ & ---  &  $ (2.8 \pm 1.6)\times 10^{-1} $  &  $ (2.3 \pm 0.2)\times 10^{0} $  &  $ (6.0 \pm 0.4)\times 10^{0} $  &  $ (1.3 \pm 0.1)\times 10^{1} $  \\ 
$ 500$ & ---  &  $ (3.5 \pm 1.7)\times 10^{-1} $  &  $ (2.0 \pm 0.2)\times 10^{0} $  &  $ (4.5 \pm 0.3)\times 10^{0} $  &  $ (1.0 \pm 0.1)\times 10^{1} $  \\ 
$ 1100$ & ---  &  $ (1.7 \pm 33.0)\times 10^{-3} $  &  $ (1.4 \pm 0.4)\times 10^{-1} $  &  $ (4.2 \pm 0.7)\times 10^{-1} $  &  $ (1.2 \pm 0.1)\times 10^{0} $  \\ 
 \hline
$\lambda $ & \multicolumn{5}{c}{${\bf  z = 2.5 }$--${\bf 3.0 }$}  \\
($\micro \rm m$) &  $  {\rm log}(M/{\rm M}_{\odot}) = 9.0$--$10.0 $  &  $  {\rm log}(M/{\rm M}_{\odot}) = 10.0$--$11.0 $  &  $  {\rm log}(M/{\rm M}_{\odot}) = 11.0$--$12.0 $  &  $  {\rm log}(M/{\rm M}_{\odot}) = 9.0$--$9.5 $  &  $  {\rm log}(M/{\rm M}_{\odot}) = 9.5$--$10.0 $  \\
 \hline
$ 24$ & ---  &  $ (9.3 \pm 16.0)\times 10^{-4} $  &  $ (2.0 \pm 0.2)\times 10^{-2} $  &  $ (7.4 \pm 0.4)\times 10^{-2} $  &  $ (2.1 \pm 0.2)\times 10^{-1} $  \\ 
$ 70$ & ---  &  ---  &  $ (1.4 \pm 1.0)\times 10^{-1} $  &  $ (2.1 \pm 1.4)\times 10^{-1} $  &  $ (3.2 \pm 2.2)\times 10^{-1} $  \\ 
$ 100$ & $ (1.5 \pm 110.0)\times 10^{-3} $  &  $ (2.1 \pm 0.7)\times 10^{-1} $  &  $ (4.2 \pm 0.9)\times 10^{-1} $  &  $ (1.1 \pm 0.1)\times 10^{0} $  &  $ (3.0 \pm 0.2)\times 10^{0} $  \\ 
$ 160$ & ---  &  $ (1.6 \pm 5.8)\times 10^{-1} $  &  $ (7.8 \pm 6.5)\times 10^{-1} $  &  $ (2.7 \pm 1.0)\times 10^{0} $  &  $ (7.2 \pm 1.4)\times 10^{0} $  \\ 
$ 250$ & ---  &  $ (2.9 \pm 1.5)\times 10^{-1} $  &  $ (1.4 \pm 0.2)\times 10^{0} $  &  $ (3.7 \pm 0.3)\times 10^{0} $  &  $ (1.1 \pm 0.1)\times 10^{1} $  \\ 
$ 350$ & ---  &  $ (4.1 \pm 1.7)\times 10^{-1} $  &  $ (1.6 \pm 0.2)\times 10^{0} $  &  $ (4.7 \pm 0.3)\times 10^{0} $  &  $ (1.4 \pm 0.1)\times 10^{1} $  \\ 
$ 500$ & ---  &  $ (3.0 \pm 1.7)\times 10^{-1} $  &  $ (1.0 \pm 0.2)\times 10^{0} $  &  $ (3.7 \pm 0.3)\times 10^{0} $  &  $ (1.1 \pm 0.1)\times 10^{1} $  \\ 
$ 1100$ & ---  &  $ (4.0 \pm 3.7)\times 10^{-2} $  &  $ (1.5 \pm 0.4)\times 10^{-1} $  &  $ (5.4 \pm 0.7)\times 10^{-1} $  &  $ (1.8 \pm 0.1)\times 10^{0} $  \\ 
 \hline
$\lambda $ & \multicolumn{5}{c}{${\bf  z = 3.0 }$--${\bf 3.5 }$}  \\
($\micro \rm m$) &  $  {\rm log}(M/{\rm M}_{\odot}) = 9.0$--$10.0 $  &  $  {\rm log}(M/{\rm M}_{\odot}) = 10.0$--$11.0 $  &  $  {\rm log}(M/{\rm M}_{\odot}) = 11.0$--$12.0 $  &  $  {\rm log}(M/{\rm M}_{\odot}) = 9.0$--$9.5 $  &  $  {\rm log}(M/{\rm M}_{\odot}) = 9.5$--$10.0 $  \\
 \hline
$ 24$ & ---  &  ---  &  $ (1.1 \pm 0.2)\times 10^{-2} $  &  $ (6.1 \pm 0.5)\times 10^{-2} $  &  $ (1.7 \pm 0.1)\times 10^{-1} $  \\ 
$ 70$ & ---  &  ---  &  ---  &  ---  &  $ (1.2 \pm 0.3)\times 10^{0} $  \\ 
$ 100$ & ---  &  $ (5.8 \pm 9.6)\times 10^{-2} $  &  $ (2.6 \pm 1.1)\times 10^{-1} $  &  $ (1.4 \pm 0.2)\times 10^{0} $  &  $ (2.8 \pm 0.3)\times 10^{0} $  \\ 
$ 160$ & ---  &  ---  &  $ (6.4 \pm 7.9)\times 10^{-1} $  &  $ (2.8 \pm 1.4)\times 10^{0} $  &  $ (6.3 \pm 2.5)\times 10^{0} $  \\ 
$ 250$ & ---  &  ---  &  $ (1.3 \pm 0.2)\times 10^{0} $  &  $ (5.1 \pm 0.4)\times 10^{0} $  &  $ (1.2 \pm 0.1)\times 10^{1} $  \\ 
$ 350$ & ---  &  $ (8.5 \pm 25.0)\times 10^{-2} $  &  $ (1.8 \pm 0.3)\times 10^{0} $  &  $ (7.1 \pm 0.5)\times 10^{0} $  &  $ (1.5 \pm 0.1)\times 10^{1} $  \\ 
$ 500$ & $ (3.1 \pm 47.0)\times 10^{-2} $  &  $ (3.1 \pm 2.6)\times 10^{-1} $  &  $ (1.8 \pm 0.3)\times 10^{0} $  &  $ (6.6 \pm 0.5)\times 10^{0} $  &  $ (1.3 \pm 0.1)\times 10^{1} $  \\ 
$ 1100$ & ---  &  ---  &  $ (1.9 \pm 0.5)\times 10^{-1} $  &  $ (7.9 \pm 1.2)\times 10^{-1} $  &  $ (2.0 \pm 0.2)\times 10^{0} $  \\ 
 \hline
$\lambda $ & \multicolumn{5}{c}{${\bf  z = 3.5 }$--${\bf 4.0 }$}  \\
($\micro \rm m$) &  $  {\rm log}(M/{\rm M}_{\odot}) = 9.0$--$10.0 $  &  $  {\rm log}(M/{\rm M}_{\odot}) = 10.0$--$11.0 $  &  $  {\rm log}(M/{\rm M}_{\odot}) = 11.0$--$12.0 $  &  $  {\rm log}(M/{\rm M}_{\odot}) = 9.0$--$9.5 $  &  $  {\rm log}(M/{\rm M}_{\odot}) = 9.5$--$10.0 $  \\
 \hline
$ 24$ & $ (6.4 \pm 0.2)\times 10^{-2} $  &  ---  &  ---  &  $ (3.1 \pm 1.0)\times 10^{-2} $  &  $ (1.1 \pm 0.3)\times 10^{-1} $  \\ 
$ 70$ & $ (7.8 \pm 1000.0)\times 10^{-1} $  &  ---  &  ---  &  $ (3.7 \pm 4.5)\times 10^{-1} $  &  $ (3.3 \pm 8.7)\times 10^{-1} $  \\ 
$ 100$ & $ (2.3 \pm 1500.0)\times 10^{-1} $  &  $ (3.7 \pm 27.0)\times 10^{-2} $  &  $ (8.2 \pm 22.0)\times 10^{-2} $  &  $ (4.7 \pm 3.7)\times 10^{-1} $  &  $ (2.9 \pm 7.1)\times 10^{-1} $  \\ 
$ 160$ & $ (3.8 \pm 1400.0)\times 10^{-1} $  &  ---  &  ---  &  $ (2.5 \pm 3.1)\times 10^{0} $  &  $ (5.4 \pm 6.6)\times 10^{0} $  \\ 
$ 250$ & $ (7.0 \pm 0.6)\times 10^{0} $  &  ---  &  $ (5.6 \pm 5.3)\times 10^{-1} $  &  $ (3.2 \pm 1.0)\times 10^{0} $  &  $ (1.1 \pm 0.2)\times 10^{1} $  \\ 
$ 350$ & $ (7.0 \pm 0.6)\times 10^{0} $  &  ---  &  $ (7.1 \pm 6.3)\times 10^{-1} $  &  $ (3.8 \pm 1.2)\times 10^{0} $  &  $ (1.6 \pm 0.3)\times 10^{1} $  \\ 
$ 500$ & $ (6.2 \pm 0.7)\times 10^{0} $  &  $ (9.9 \pm 7.1)\times 10^{-1} $  &  $ (1.0 \pm 0.6)\times 10^{0} $  &  $ (3.3 \pm 1.1)\times 10^{0} $  &  $ (1.5 \pm 0.2)\times 10^{1} $  \\ 
$ 1100$ & $ (9.0 \pm 0.5)\times 10^{-1} $  &  $ (1.2 \pm 1.4)\times 10^{-1} $  &  $ (2.1 \pm 1.3)\times 10^{-1} $  &  $ (5.8 \pm 2.5)\times 10^{-1} $  &  $ (2.4 \pm 0.0)\times 10^{0} $  \\ 
\hline
\end{tabular}
\tablecomments{Stacked Flux Densities in $\rm mJy$. Errors estimated with a extended bootstrap technique described in \S~\ref{sec:errors}.  Blank spaces represent where the algorithm fails to adequately converge.  Flux densities, converted to $\nu I_{\nu}$, are shown in Figure~\ref{fig:bethcib}. }
\label{tab:stacked_sf}
\end{table*}

%% file: t2.tex
\begin{table*}
\tiny
 \centering
\caption{Stacked Flux Densities of Quiescent Galaxies}
 \begin{tabular}{cccc}
 \hline
$\lambda $ & \multicolumn{3}{c}{${\bf  z = 0.0 }$--${\bf 0.5 }$}  \\
($\micro \rm m$) &  $  {\rm log}(M/{\rm M}_{\odot}) = 9.0$--$10.0 $  &  $  {\rm log}(M/{\rm M}_{\odot}) = 10.0$--$11.0 $  &  $  {\rm log}(M/{\rm M}_{\odot}) = 11.0$--$12.0 $  \\
 \hline
$ 24$ & $ (1.4 \pm 0.5)\times 10^{-2} $  &  $ (2.8 \pm 0.4)\times 10^{-2} $  &  $ (1.0 \pm 0.2)\times 10^{-1} $  \\ 
$ 70$ & $ (8.3 \pm 20.0)\times 10^{-2} $  &  $ (1.5 \pm 1.9)\times 10^{-1} $  &  ---  \\ 
$ 100$ & $ (8.2 \pm 1.7)\times 10^{-1} $  &  $ (7.4 \pm 1.5)\times 10^{-1} $  &  $ (1.7 \pm 0.6)\times 10^{0} $  \\ 
$ 160$ & $ (4.7 \pm 13.0)\times 10^{-1} $  &  $ (8.5 \pm 12.0)\times 10^{-1} $  &  $ (4.0 \pm 5.5)\times 10^{0} $  \\ 
$ 250$ & ---  &  $ (5.6 \pm 2.8)\times 10^{-1} $  &  $ (2.4 \pm 1.1)\times 10^{0} $  \\ 
$ 350$ & ---  &  $ (2.6 \pm 3.0)\times 10^{-1} $  &  $ (1.3 \pm 1.3)\times 10^{0} $  \\ 
$ 500$ & ---  &  ---  &  $ (7.7 \pm 13.0)\times 10^{-1} $  \\ 
$ 1100$ & ---  &  ---  &  ---  \\ 
 \hline
$\lambda $ & \multicolumn{3}{c}{${\bf  z = 0.5 }$--${\bf 1.0 }$}  \\
($\micro \rm m$) &  $  {\rm log}(M/{\rm M}_{\odot}) = 9.0$--$10.0 $  &  $  {\rm log}(M/{\rm M}_{\odot}) = 10.0$--$11.0 $  &  $  {\rm log}(M/{\rm M}_{\odot}) = 11.0$--$12.0 $  \\
 \hline
$ 24$ & ---  &  $ (1.4 \pm 0.2)\times 10^{-2} $  &  $ (2.4 \pm 0.7)\times 10^{-2} $  \\ 
$ 70$ & $ (6.5 \pm 11.0)\times 10^{-2} $  &  ---  &  ---  \\ 
$ 100$ & $ (7.6 \pm 8.5)\times 10^{-2} $  &  $ (3.3 \pm 0.7)\times 10^{-1} $  &  $ (3.8 \pm 2.2)\times 10^{-1} $  \\ 
$ 160$ & $ (4.0 \pm 6.3)\times 10^{-1} $  &  $ (5.7 \pm 51.0)\times 10^{-2} $  &  ---  \\ 
$ 250$ & $ (3.8 \pm 1.9)\times 10^{-1} $  &  ---  &  $ (4.7 \pm 5.1)\times 10^{-1} $  \\ 
$ 350$ & $ (5.7 \pm 2.1)\times 10^{-1} $  &  ---  &  $ (2.1 \pm 6.0)\times 10^{-1} $  \\ 
$ 500$ & $ (3.6 \pm 1.9)\times 10^{-1} $  &  ---  &  ---  \\ 
$ 1100$ & $ (4.5 \pm 3.6)\times 10^{-2} $  &  ---  &  $ (6.0 \pm 11.0)\times 10^{-2} $  \\ 
 \hline
$\lambda $ & \multicolumn{3}{c}{${\bf  z = 1.0 }$--${\bf 1.5 }$}  \\
($\micro \rm m$) &  $  {\rm log}(M/{\rm M}_{\odot}) = 9.0$--$10.0 $  &  $  {\rm log}(M/{\rm M}_{\odot}) = 10.0$--$11.0 $  &  $  {\rm log}(M/{\rm M}_{\odot}) = 11.0$--$12.0 $  \\
 \hline
$ 24$ & ---  &  $ (5.9 \pm 1.7)\times 10^{-3} $  &  $ (2.1 \pm 0.6)\times 10^{-2} $  \\ 
$ 70$ & ---  &  ---  &  $ (2.4 \pm 2.2)\times 10^{-1} $  \\ 
$ 100$ & ---  &  ---  &  ---  \\ 
$ 160$ & ---  &  $ (3.2 \pm 4.7)\times 10^{-1} $  &  $ (6.1 \pm 14.0)\times 10^{-1} $  \\ 
$ 250$ & ---  &  $ (2.4 \pm 1.4)\times 10^{-1} $  &  $ (1.1 \pm 0.4)\times 10^{0} $  \\ 
$ 350$ & ---  &  $ (3.1 \pm 1.7)\times 10^{-1} $  &  $ (1.7 \pm 0.5)\times 10^{0} $  \\ 
$ 500$ & ---  &  $ (1.2 \pm 1.6)\times 10^{-1} $  &  $ (1.4 \pm 0.5)\times 10^{0} $  \\ 
$ 1100$ & ---  &  ---  &  ---  \\ 
 \hline
$\lambda $ & \multicolumn{3}{c}{${\bf  z = 1.5 }$--${\bf 2.0 }$}  \\
($\micro \rm m$) &  $  {\rm log}(M/{\rm M}_{\odot}) = 9.0$--$10.0 $  &  $  {\rm log}(M/{\rm M}_{\odot}) = 10.0$--$11.0 $  &  $  {\rm log}(M/{\rm M}_{\odot}) = 11.0$--$12.0 $  \\
 \hline
$ 24$ & ---  &  $ (1.3 \pm 1.8)\times 10^{-3} $  &  $ (1.1 \pm 0.7)\times 10^{-2} $  \\ 
$ 70$ & ---  &  ---  &  ---  \\ 
$ 100$ & ---  &  ---  &  $ (2.1 \pm 24.0)\times 10^{-2} $  \\ 
$ 160$ & ---  &  ---  &  $ (9.2 \pm 22.0)\times 10^{-1} $  \\ 
$ 250$ & ---  &  $ (1.6 \pm 1.9)\times 10^{-1} $  &  ---  \\ 
$ 350$ & $ (2.3 \pm 3.4)\times 10^{-1} $  &  $ (3.9 \pm 2.4)\times 10^{-1} $  &  $ (4.7 \pm 7.4)\times 10^{-1} $  \\ 
$ 500$ & $ (3.8 \pm 3.6)\times 10^{-1} $  &  $ (3.7 \pm 2.3)\times 10^{-1} $  &  $ (4.9 \pm 7.6)\times 10^{-1} $  \\ 
$ 1100$ & ---  &  ---  &  $ (1.6 \pm 1.2)\times 10^{-1} $  \\ 
 \hline
$\lambda $ & \multicolumn{3}{c}{${\bf  z = 2.0 }$--${\bf 2.5 }$}  \\
($\micro \rm m$) &  $  {\rm log}(M/{\rm M}_{\odot}) = 9.0$--$10.0 $  &  $  {\rm log}(M/{\rm M}_{\odot}) = 10.0$--$11.0 $  &  $  {\rm log}(M/{\rm M}_{\odot}) = 11.0$--$12.0 $  \\
 \hline
$ 24$ & ---  &  $ (8.7 \pm 3.4)\times 10^{-3} $  &  $ (5.7 \pm 0.7)\times 10^{-2} $  \\ 
$ 70$ & $ (2.5 \pm 32.0)\times 10^{-2} $  &  ---  &  $ (3.1 \pm 2.8)\times 10^{-1} $  \\ 
$ 100$ & ---  &  ---  &  $ (1.4 \pm 0.2)\times 10^{0} $  \\ 
$ 160$ & $ (3.7 \pm 2.2)\times 10^{0} $  &  $ (8.3 \pm 12.0)\times 10^{-1} $  &  $ (4.1 \pm 1.9)\times 10^{0} $  \\ 
$ 250$ & $ (9.3 \pm 49.0)\times 10^{-2} $  &  $ (7.1 \pm 3.1)\times 10^{-1} $  &  $ (4.1 \pm 0.6)\times 10^{0} $  \\ 
$ 350$ & $ (1.7 \pm 0.6)\times 10^{0} $  &  $ (1.0 \pm 0.4)\times 10^{0} $  &  $ (4.6 \pm 0.7)\times 10^{0} $  \\ 
$ 500$ & $ (1.9 \pm 0.6)\times 10^{0} $  &  $ (8.2 \pm 3.5)\times 10^{-1} $  &  $ (3.6 \pm 0.6)\times 10^{0} $  \\ 
$ 1100$ & ---  &  ---  &  $ (2.3 \pm 1.1)\times 10^{-1} $  \\ 
 \hline
$\lambda $ & \multicolumn{3}{c}{${\bf  z = 2.5 }$--${\bf 3.0 }$}  \\
($\micro \rm m$) &  $  {\rm log}(M/{\rm M}_{\odot}) = 9.0$--$10.0 $  &  $  {\rm log}(M/{\rm M}_{\odot}) = 10.0$--$11.0 $  &  $  {\rm log}(M/{\rm M}_{\odot}) = 11.0$--$12.0 $  \\
 \hline
$ 24$ & ---  &  $ (2.0 \pm 0.3)\times 10^{-2} $  &  $ (8.2 \pm 0.5)\times 10^{-2} $  \\ 
$ 70$ & ---  &  ---  &  $ (8.0 \pm 24.0)\times 10^{-2} $  \\ 
$ 100$ & ---  &  $ (2.7 \pm 1.5)\times 10^{-1} $  &  $ (7.7 \pm 1.7)\times 10^{-1} $  \\ 
$ 160$ & ---  &  $ (5.1 \pm 15.0)\times 10^{-1} $  &  $ (1.5 \pm 1.6)\times 10^{0} $  \\ 
$ 250$ & ---  &  $ (1.1 \pm 0.3)\times 10^{0} $  &  $ (3.8 \pm 0.4)\times 10^{0} $  \\ 
$ 350$ & ---  &  $ (1.8 \pm 0.4)\times 10^{0} $  &  $ (4.1 \pm 0.6)\times 10^{0} $  \\ 
$ 500$ & ---  &  $ (1.8 \pm 0.4)\times 10^{0} $  &  $ (3.2 \pm 0.5)\times 10^{0} $  \\ 
$ 1100$ & ---  &  $ (2.0 \pm 0.8)\times 10^{-1} $  &  $ (3.3 \pm 1.1)\times 10^{-1} $  \\ 
 \hline
$\lambda $ & \multicolumn{3}{c}{${\bf  z = 3.0 }$--${\bf 3.5 }$}  \\
($\micro \rm m$) &  $  {\rm log}(M/{\rm M}_{\odot}) = 9.0$--$10.0 $  &  $  {\rm log}(M/{\rm M}_{\odot}) = 10.0$--$11.0 $  &  $  {\rm log}(M/{\rm M}_{\odot}) = 11.0$--$12.0 $  \\
 \hline
$ 24$ & ---  &  $ (2.0 \pm 0.4)\times 10^{-2} $  &  $ (8.5 \pm 0.7)\times 10^{-2} $  \\ 
$ 70$ & $ (2.4 \pm 200.0)\times 10^{0} $  &  ---  &  ---  \\ 
$ 100$ & $ (4.2 \pm 280.0)\times 10^{0} $  &  $ (7.1 \pm 2.2)\times 10^{-1} $  &  $ (1.7 \pm 0.2)\times 10^{0} $  \\ 
$ 160$ & ---  &  $ (2.6 \pm 16.0)\times 10^{-1} $  &  $ (3.2 \pm 1.9)\times 10^{0} $  \\ 
$ 250$ & $ (2.3 \pm 1.1)\times 10^{0} $  &  $ (9.9 \pm 4.4)\times 10^{-1} $  &  $ (5.7 \pm 0.6)\times 10^{0} $  \\ 
$ 350$ & $ (2.5 \pm 1.1)\times 10^{0} $  &  $ (1.6 \pm 0.6)\times 10^{0} $  &  $ (6.2 \pm 0.8)\times 10^{0} $  \\ 
$ 500$ & $ (4.8 \pm 1.3)\times 10^{0} $  &  $ (1.2 \pm 0.5)\times 10^{0} $  &  $ (4.8 \pm 0.8)\times 10^{0} $  \\ 
$ 1100$ & ---  &  $ (2.3 \pm 1.1)\times 10^{-1} $  &  $ (4.2 \pm 1.4)\times 10^{-1} $  \\ 
 \hline
$\lambda $ & \multicolumn{3}{c}{${\bf  z = 3.5 }$--${\bf 4.0 }$}  \\
($\micro \rm m$) &  $  {\rm log}(M/{\rm M}_{\odot}) = 9.0$--$10.0 $  &  $  {\rm log}(M/{\rm M}_{\odot}) = 10.0$--$11.0 $  &  $  {\rm log}(M/{\rm M}_{\odot}) = 11.0$--$12.0 $  \\
 \hline
$ 24$ & ---  &  $ (4.2 \pm 5.1)\times 10^{-3} $  &  $ (5.4 \pm 1.1)\times 10^{-2} $  \\ 
$ 70$ & ---  &  $ (1.6 \pm 4.6)\times 10^{-1} $  &  ---  \\ 
$ 100$ & ---  &  ---  &  $ (6.6 \pm 4.8)\times 10^{-1} $  \\ 
$ 160$ & ---  &  $ (1.5 \pm 290.0)\times 10^{-2} $  &  $ (5.1 \pm 36.0)\times 10^{-1} $  \\ 
$ 250$ & ---  &  $ (3.7 \pm 0.8)\times 10^{0} $  &  $ (3.8 \pm 1.1)\times 10^{0} $  \\ 
$ 350$ & ---  &  $ (4.7 \pm 1.1)\times 10^{0} $  &  $ (5.9 \pm 1.4)\times 10^{0} $  \\ 
$ 500$ & ---  &  $ (3.2 \pm 1.1)\times 10^{0} $  &  $ (5.9 \pm 1.5)\times 10^{0} $  \\ 
$ 1100$ & ---  &  $ (4.3 \pm 3.4)\times 10^{-1} $  &  $ (1.7 \pm 0.4)\times 10^{0} $  \\ 
\hline
\end{tabular}
\tablecomments{Stacked Flux Densities in $\rm mJy$. Errors estimated with a extended bootstrap technique described in \S~\ref{sec:errors}.  Blank spaces represent where the algorithm fails to adequately converge.  Flux densities, converted to $\nu I_{\nu}$, are shown in Figure~\ref{fig:bethcib}. }
\label{tab:stacked_passive}
\end{table*}

%% file: t3.tex
\begin{table*}
 \scriptsize 
 \centering
\caption{Stacked Intensities in Bins of Redshift}
 \begin{tabular}{cccccc}
 \hline
 \hline
 Band &  Type & $ z = 0.0$--$1.0 $  & $ z = 1.0$--$2.0 $  & $ z = 2.0$--$3.0 $  & $ z = 3.0$--$4.0 $  \\ 
 ($\micro m$) &  $ $ & ($\rm nW\, m^{-2}\, sr^{-1}$) & ($\rm nW\, m^{-2}\, sr^{-1}$) & ($\rm nW\, m^{-2}\, sr^{-1}$) & ($\rm nW\, m^{-2}\, sr^{-1}$) \\ 
 \hline
\multirow{3}{*}{24} &  Stack &  $ (9.2 \pm 0.2)\times 10^{-1} $  $[50 \%]$ &  $ (6.8 \pm 0.2)\times 10^{-1} $  $[36 \%]$ &  $ (2.1 \pm 0.1)\times 10^{-1} $  $[11 \%]$ &  $ (3.0 \pm 0.2)\times 10^{-2} $  $[1 \%]$ \\
& CC &  $ (9.2 \pm 0.2)\times 10^{-1} $  $[39 \%]$ &  $ (7.3 \pm 0.3)\times 10^{-1} $  $[31 \%]$ &  $ (3.6 \pm 0.4)\times 10^{-1} $  $[15 \%]$ &  $ (3.0 \pm 0.5)\times 10^{-1} $  $[13 \%]$ \\
& SED &  $ (1.0 \pm 0.4)\times 10^{0} $  $[44 \%]$ &  $ (8.3 \pm 2.9)\times 10^{-1} $  $[35 \%]$ &  $ (3.5 \pm 1.3)\times 10^{-1} $  $[15 \%]$ &  $ (3.6 \pm 0.0)\times 10^{-1} $  $[15 \%]$ \\
 \hline
\multirow{3}{*}{70} &  Stack &  $ (2.2 \pm 0.1)\times 10^{0} $  $[66 \%]$ &  $ (8.7 \pm 1.1)\times 10^{-1} $  $[26 \%]$ &  $ (2.1 \pm 0.7)\times 10^{-1} $  $[6 \%]$ &  $ (2.7 \pm 3.7)\times 10^{-2} $  $[0 \%]$ \\
& CC &  $ (2.2 \pm 0.1)\times 10^{0} $  $[62 \%]$ &  $ (8.7 \pm 2.0)\times 10^{-1} $  $[24 \%]$ &  $ (4.3 \pm 8.0)\times 10^{-1} $  $[12 \%]$ &  $ (5.0 \pm 130.0)\times 10^{-2} $  $[1 \%]$ \\
& SED &  $ (2.7 \pm 1.0)\times 10^{0} $  $[76 \%]$ &  $ (2.1 \pm 0.7)\times 10^{0} $  $[59 \%]$ &  $ (8.7 \pm 3.2)\times 10^{-1} $  $[24 \%]$ &  $ (9.3 \pm 0.0)\times 10^{-1} $  $[26 \%]$ \\
 \hline
\multirow{3}{*}{100} &  Stack &  $ (4.9 \pm 0.2)\times 10^{0} $  $[56 \%]$ &  $ (3.0 \pm 0.1)\times 10^{0} $  $[33 \%]$ &  $ (6.9 \pm 0.4)\times 10^{-1} $  $[7 \%]$ &  $ (1.5 \pm 0.2)\times 10^{-1} $  $[1 \%]$ \\
& CC &  $ (4.9 \pm 0.2)\times 10^{0} $  $[44 \%]$ &  $ (3.2 \pm 0.2)\times 10^{0} $  $[29 \%]$ &  $ (1.2 \pm 0.4)\times 10^{0} $  $[10 \%]$ &  $ (1.7 \pm 0.6)\times 10^{0} $  $[15 \%]$ \\
& SED &  $ (3.9 \pm 1.4)\times 10^{0} $  $[35 \%]$ &  $ (3.0 \pm 1.0)\times 10^{0} $  $[26 \%]$ &  $ (1.2 \pm 0.5)\times 10^{0} $  $[10 \%]$ &  $ (1.3 \pm 0.0)\times 10^{0} $  $[11 \%]$ \\
 \hline
\multirow{3}{*}{160} &  Stack &  $ (4.8 \pm 0.3)\times 10^{0} $  $[51 \%]$ &  $ (3.4 \pm 0.3)\times 10^{0} $  $[36 \%]$ &  $ (1.0 \pm 0.2)\times 10^{0} $  $[10 \%]$ &  $ (1.8 \pm 0.9)\times 10^{-1} $  $[1 \%]$ \\
& CC &  $ (4.8 \pm 0.3)\times 10^{0} $  $[37 \%]$ &  $ (3.7 \pm 0.6)\times 10^{0} $  $[28 \%]$ &  $ (1.8 \pm 2.2)\times 10^{0} $  $[13 \%]$ &  $ (2.5 \pm 2.8)\times 10^{0} $  $[19 \%]$ \\
& SED &  $ (5.3 \pm 2.0)\times 10^{0} $  $[41 \%]$ &  $ (4.6 \pm 1.6)\times 10^{0} $  $[36 \%]$ &  $ (1.9 \pm 0.7)\times 10^{0} $  $[14 \%]$ &  $ (2.1 \pm 0.0)\times 10^{0} $  $[16 \%]$ \\
 \hline
\multirow{3}{*}{250} &  Stack &  $ (2.7 \pm 0.1)\times 10^{0} $  $[38 \%]$ &  $ (3.0 \pm 0.2)\times 10^{0} $  $[43 \%]$ &  $ (1.0 \pm 0.1)\times 10^{0} $  $[14 \%]$ &  $ (2.5 \pm 0.2)\times 10^{-1} $  $[3 \%]$ \\
& CC &  $ (2.7 \pm 0.1)\times 10^{0} $  $[23 \%]$ &  $ (3.3 \pm 0.2)\times 10^{0} $  $[29 \%]$ &  $ (1.9 \pm 0.4)\times 10^{0} $  $[16 \%]$ &  $ (3.5 \pm 0.5)\times 10^{0} $  $[30 \%]$ \\
& SED &  $ (2.9 \pm 1.1)\times 10^{0} $  $[25 \%]$ &  $ (4.3 \pm 1.5)\times 10^{0} $  $[37 \%]$ &  $ (2.6 \pm 0.9)\times 10^{0} $  $[22 \%]$ &  $ (3.1 \pm 0.0)\times 10^{0} $  $[27 \%]$ \\
 \hline
\multirow{3}{*}{350} &  Stack &  $ (1.3 \pm 0.1)\times 10^{0} $  $[29 \%]$ &  $ (2.0 \pm 0.1)\times 10^{0} $  $[45 \%]$ &  $ (8.8 \pm 0.4)\times 10^{-1} $  $[20 \%]$ &  $ (2.4 \pm 0.2)\times 10^{-1} $  $[5 \%]$ \\
& CC &  $ (1.3 \pm 0.1)\times 10^{0} $  $[14 \%]$ &  $ (2.2 \pm 0.1)\times 10^{0} $  $[24 \%]$ &  $ (2.0 \pm 0.3)\times 10^{0} $  $[22 \%]$ &  $ (3.4 \pm 0.5)\times 10^{0} $  $[38 \%]$ \\
& SED &  $ (1.1 \pm 0.4)\times 10^{0} $  $[12 \%]$ &  $ (2.3 \pm 0.8)\times 10^{0} $  $[26 \%]$ &  $ (2.1 \pm 0.7)\times 10^{0} $  $[23 \%]$ &  $ (3.1 \pm 0.0)\times 10^{0} $  $[35 \%]$ \\
 \hline
\multirow{3}{*}{500} &  Stack &  $ (4.1 \pm 0.3)\times 10^{-1} $  $[21 \%]$ &  $ (8.0 \pm 0.5)\times 10^{-1} $  $[43 \%]$ &  $ (4.8 \pm 0.3)\times 10^{-1} $  $[25 \%]$ &  $ (1.6 \pm 0.1)\times 10^{-1} $  $[8 \%]$ \\
& CC &  $ (4.1 \pm 0.3)\times 10^{-1} $  $[8 \%]$ &  $ (8.9 \pm 0.7)\times 10^{-1} $  $[18 \%]$ &  $ (1.2 \pm 0.2)\times 10^{0} $  $[25 \%]$ &  $ (2.3 \pm 0.3)\times 10^{0} $  $[47 \%]$ \\
& SED &  $ (3.3 \pm 1.3)\times 10^{-1} $  $[6 \%]$ &  $ (8.2 \pm 2.7)\times 10^{-1} $  $[17 \%]$ &  $ (1.0 \pm 0.3)\times 10^{0} $  $[21 \%]$ &  $ (1.7 \pm 0.0)\times 10^{0} $  $[36 \%]$ \\
 \hline
\multirow{3}{*}{1100} &  Stack &  $ (9.6 \pm 1.5)\times 10^{-3} $  $[14 \%]$ &  $ (2.5 \pm 0.3)\times 10^{-2} $  $[38 \%]$ &  $ (2.0 \pm 0.2)\times 10^{-2} $  $[31 \%]$ &  $ (9.6 \pm 1.0)\times 10^{-3} $  $[14 \%]$ \\
& CC &  $ (9.6 \pm 1.5)\times 10^{-3} $  $[5 \%]$ &  $ (2.6 \pm 0.4)\times 10^{-2} $  $[14 \%]$ &  $ (3.5 \pm 1.7)\times 10^{-2} $  $[20 \%]$ &  $ (1.0 \pm 0.3)\times 10^{-1} $  $[59 \%]$ \\
& SED &  $ (1.2 \pm 0.5)\times 10^{-2} $  $[6 \%]$ &  $ (4.0 \pm 1.3)\times 10^{-2} $  $[22 \%]$ &  $ (6.3 \pm 1.9)\times 10^{-2} $  $[36 \%]$ &  $ (1.2 \pm 0.0)\times 10^{-1} $  $[70 \%]$ \\
 \hline
\end{tabular}
\tablecomments{Stacked intensities in units of $\rm nW\, m^{-2}\, sr^{-1}$. Uncorrected intensities are labeled \lq\lq Stack\rq\rq, completeness-corrected values are labeled \lq\lq CC\rq\rq, and the best-fit SED values are labeled \lq\lq SED\rq\rq.  Errors estimated with a Monte Carlo simulation described in \S~\ref{sec:errors}.   Blank spaces represent where the algorithm fails to adequately converge.}
\label{tab:cib_redshifts}
\end{table*}

%% file: t4.tex
\begin{table*}
\tiny
 \centering
\caption{Stacked Intensities of Star-Forming Galaxie in Bins of Stellar Mass}
 \begin{tabular}{cccccccccc}
 \hline
 \hline
 Band &  Type & $ {\rm log}(M/{\rm M}_{\odot}) = 9.0$--$9.5 $  & $ {\rm log}(M/{\rm M}_{\odot}) = 9.5$--$10.0 $  & $ {\rm log}(M/{\rm M}_{\odot}) = 10.0$--$10.5 $  & $ {\rm log}(M/{\rm M}_{\odot}) = 10.5$--$11.0 $  & $ {\rm log}(M/{\rm M}_{\odot}) = 11.0$--$12.0 $  \\ 
 ($\micro m$) &  $ $ & ($\rm nW\, m^{-2}\, sr^{-1}$) & ($\rm nW\, m^{-2}\, sr^{-1}$) & ($\rm nW\, m^{-2}\, sr^{-1}$) & ($\rm nW\, m^{-2}\, sr^{-1}$) & ($\rm nW\, m^{-2}\, sr^{-1}$) \\ 
 \hline
\multirow{3}{*}{24} &  Stack &  $ (1.2 \pm 0.1)\times 10^{-1} $  $[6 \%]$ &  $ (4.0 \pm 0.1)\times 10^{-1} $  $[21 \%]$ &  $ (6.9 \pm 0.2)\times 10^{-1} $  $[37 \%]$ &  $ (4.6 \pm 0.1)\times 10^{-1} $  $[24 \%]$ &  $ (1.1 \pm 0.0)\times 10^{-1} $  $[6 \%]$ \\
& CC &  $ (1.2 \pm 0.1)\times 10^{-1} $  $[5 \%]$ &  $ (4.8 \pm 0.4)\times 10^{-1} $  $[20 \%]$ &  $ (1.1 \pm 0.1)\times 10^{0} $  $[45 \%]$ &  $ (4.7 \pm 0.1)\times 10^{-1} $  $[20 \%]$ &  $ (1.1 \pm 0.0)\times 10^{-1} $  $[4 \%]$ \\
& SED &  $ (2.0 \pm 0.9)\times 10^{-1} $  $[8 \%]$ &  $ (5.7 \pm 2.5)\times 10^{-1} $  $[24 \%]$ &  $ (1.1 \pm 0.4)\times 10^{0} $  $[49 \%]$ &  $ (4.8 \pm 2.0)\times 10^{-1} $  $[20 \%]$ &  $ (9.6 \pm 4.0)\times 10^{-2} $  $[4 \%]$ \\
\multirow{3}{*}{70} &  Stack &  $ (2.0 \pm 1.2)\times 10^{-1} $  $[6 \%]$ &  $ (9.0 \pm 1.2)\times 10^{-1} $  $[27 \%]$ &  $ (1.3 \pm 0.1)\times 10^{0} $  $[38 \%]$ &  $ (7.1 \pm 0.6)\times 10^{-1} $  $[21 \%]$ &  $ (1.7 \pm 0.2)\times 10^{-1} $  $[5 \%]$ \\
& CC &  $ (2.0 \pm 1.9)\times 10^{-1} $  $[5 \%]$ &  $ (9.0 \pm 7.9)\times 10^{-1} $  $[25 \%]$ &  $ (1.5 \pm 1.3)\times 10^{0} $  $[41 \%]$ &  $ (7.2 \pm 0.7)\times 10^{-1} $  $[20 \%]$ &  $ (1.7 \pm 0.2)\times 10^{-1} $  $[4 \%]$ \\
& SED &  $ (5.2 \pm 2.5)\times 10^{-1} $  $[14 \%]$ &  $ (1.5 \pm 0.7)\times 10^{0} $  $[43 \%]$ &  $ (3.0 \pm 1.2)\times 10^{0} $  $[83 \%]$ &  $ (1.2 \pm 0.5)\times 10^{0} $  $[32 \%]$ &  $ (2.3 \pm 0.9)\times 10^{-1} $  $[6 \%]$ \\
\multirow{3}{*}{100} &  Stack &  $ (8.8 \pm 0.8)\times 10^{-1} $  $[10 \%]$ &  $ (2.2 \pm 0.1)\times 10^{0} $  $[25 \%]$ &  $ (3.3 \pm 0.1)\times 10^{0} $  $[37 \%]$ &  $ (1.8 \pm 0.1)\times 10^{0} $  $[20 \%]$ &  $ (3.3 \pm 0.2)\times 10^{-1} $  $[3 \%]$ \\
& CC &  $ (9.7 \pm 1.2)\times 10^{-1} $  $[8 \%]$ &  $ (2.4 \pm 0.4)\times 10^{0} $  $[21 \%]$ &  $ (5.2 \pm 0.6)\times 10^{0} $  $[46 \%]$ &  $ (1.9 \pm 0.1)\times 10^{0} $  $[16 \%]$ &  $ (3.4 \pm 0.2)\times 10^{-1} $  $[3 \%]$ \\
& SED &  $ (7.4 \pm 3.6)\times 10^{-1} $  $[6 \%]$ &  $ (2.2 \pm 1.0)\times 10^{0} $  $[20 \%]$ &  $ (4.2 \pm 1.7)\times 10^{0} $  $[38 \%]$ &  $ (1.6 \pm 0.7)\times 10^{0} $  $[14 \%]$ &  $ (3.2 \pm 1.3)\times 10^{-1} $  $[2 \%]$ \\
\multirow{3}{*}{160} &  Stack &  $ (6.5 \pm 3.2)\times 10^{-1} $  $[6 \%]$ &  $ (2.1 \pm 0.3)\times 10^{0} $  $[21 \%]$ &  $ (3.6 \pm 0.3)\times 10^{0} $  $[37 \%]$ &  $ (2.3 \pm 0.2)\times 10^{0} $  $[24 \%]$ &  $ (5.1 \pm 0.5)\times 10^{-1} $  $[5 \%]$ \\
& CC &  $ (6.6 \pm 5.1)\times 10^{-1} $  $[5 \%]$ &  $ (2.3 \pm 2.3)\times 10^{0} $  $[18 \%]$ &  $ (6.4 \pm 2.9)\times 10^{0} $  $[49 \%]$ &  $ (2.4 \pm 0.2)\times 10^{0} $  $[18 \%]$ &  $ (5.2 \pm 0.5)\times 10^{-1} $  $[4 \%]$ \\
& SED &  $ (1.0 \pm 0.5)\times 10^{0} $  $[7 \%]$ &  $ (3.1 \pm 1.4)\times 10^{0} $  $[24 \%]$ &  $ (6.4 \pm 2.5)\times 10^{0} $  $[50 \%]$ &  $ (2.5 \pm 1.1)\times 10^{0} $  $[19 \%]$ &  $ (4.9 \pm 2.0)\times 10^{-1} $  $[3 \%]$ \\
\multirow{3}{*}{250} &  Stack &  $ (3.8 \pm 0.6)\times 10^{-1} $  $[5 \%]$ &  $ (1.4 \pm 0.1)\times 10^{0} $  $[20 \%]$ &  $ (2.5 \pm 0.1)\times 10^{0} $  $[35 \%]$ &  $ (1.9 \pm 0.1)\times 10^{0} $  $[27 \%]$ &  $ (5.3 \pm 0.2)\times 10^{-1} $  $[7 \%]$ \\
& CC &  $ (4.1 \pm 0.9)\times 10^{-1} $  $[3 \%]$ &  $ (1.9 \pm 0.4)\times 10^{0} $  $[16 \%]$ &  $ (6.1 \pm 0.6)\times 10^{0} $  $[53 \%]$ &  $ (2.0 \pm 0.1)\times 10^{0} $  $[17 \%]$ &  $ (5.3 \pm 0.2)\times 10^{-1} $  $[4 \%]$ \\
& SED &  $ (6.4 \pm 2.7)\times 10^{-1} $  $[5 \%]$ &  $ (2.3 \pm 0.9)\times 10^{0} $  $[19 \%]$ &  $ (6.4 \pm 2.9)\times 10^{0} $  $[56 \%]$ &  $ (2.5 \pm 1.0)\times 10^{0} $  $[21 \%]$ &  $ (6.2 \pm 2.6)\times 10^{-1} $  $[5 \%]$ \\
\multirow{3}{*}{350} &  Stack &  $ (1.8 \pm 0.4)\times 10^{-1} $  $[4 \%]$ &  $ (8.0 \pm 0.6)\times 10^{-1} $  $[18 \%]$ &  $ (1.5 \pm 0.1)\times 10^{0} $  $[33 \%]$ &  $ (1.3 \pm 0.1)\times 10^{0} $  $[29 \%]$ &  $ (4.0 \pm 0.1)\times 10^{-1} $  $[9 \%]$ \\
& CC &  $ (1.8 \pm 0.7)\times 10^{-1} $  $[2 \%]$ &  $ (1.5 \pm 0.3)\times 10^{0} $  $[16 \%]$ &  $ (4.9 \pm 0.5)\times 10^{0} $  $[55 \%]$ &  $ (1.4 \pm 0.1)\times 10^{0} $  $[15 \%]$ &  $ (4.1 \pm 0.1)\times 10^{-1} $  $[4 \%]$ \\
& SED &  $ (3.0 \pm 1.2)\times 10^{-1} $  $[3 \%]$ &  $ (1.3 \pm 0.5)\times 10^{0} $  $[14 \%]$ &  $ (4.7 \pm 2.6)\times 10^{0} $  $[53 \%]$ &  $ (1.5 \pm 0.6)\times 10^{0} $  $[17 \%]$ &  $ (4.6 \pm 1.9)\times 10^{-1} $  $[5 \%]$ \\
\multirow{3}{*}{500} &  Stack &  $ (6.3 \pm 2.8)\times 10^{-2} $  $[3 \%]$ &  $ (2.9 \pm 0.3)\times 10^{-1} $  $[15 \%]$ &  $ (5.9 \pm 0.3)\times 10^{-1} $  $[31 \%]$ &  $ (5.6 \pm 0.2)\times 10^{-1} $  $[30 \%]$ &  $ (2.0 \pm 0.1)\times 10^{-1} $  $[10 \%]$ \\
& CC &  $ (6.4 \pm 4.4)\times 10^{-2} $  $[1 \%]$ &  $ (7.9 \pm 2.1)\times 10^{-1} $  $[16 \%]$ &  $ (2.8 \pm 0.3)\times 10^{0} $  $[58 \%]$ &  $ (6.3 \pm 0.3)\times 10^{-1} $  $[13 \%]$ &  $ (2.1 \pm 0.1)\times 10^{-1} $  $[4 \%]$ \\
& SED &  $ (1.0 \pm 0.4)\times 10^{-1} $  $[2 \%]$ &  $ (5.4 \pm 2.0)\times 10^{-1} $  $[11 \%]$ &  $ (2.2 \pm 1.4)\times 10^{0} $  $[46 \%]$ &  $ (6.1 \pm 2.4)\times 10^{-1} $  $[12 \%]$ &  $ (2.1 \pm 0.9)\times 10^{-1} $  $[4 \%]$ \\
\multirow{3}{*}{1100} &  Stack &  $ (7.2 \pm 210.0)\times 10^{-5} $  $[0 \%]$ &  $ (4.6 \pm 2.4)\times 10^{-3} $  $[7 \%]$ &  $ (2.0 \pm 0.2)\times 10^{-2} $  $[31 \%]$ &  $ (2.3 \pm 0.2)\times 10^{-2} $  $[36 \%]$ &  $ (1.2 \pm 0.1)\times 10^{-2} $  $[17 \%]$ \\
& CC &  $ (7.2 \pm 340.0)\times 10^{-5} $  $[0 \%]$ &  $ (6.5 \pm 17.0)\times 10^{-3} $  $[3 \%]$ &  $ (1.2 \pm 0.3)\times 10^{-1} $  $[68 \%]$ &  $ (2.7 \pm 0.2)\times 10^{-2} $  $[15 \%]$ &  $ (1.2 \pm 0.1)\times 10^{-2} $  $[6 \%]$ \\
& SED &  $ (4.5 \pm 1.7)\times 10^{-3} $  $[2 \%]$ &  $ (3.2 \pm 1.3)\times 10^{-2} $  $[18 \%]$ &  $ (1.4 \pm 0.9)\times 10^{-1} $  $[78 \%]$ &  $ (3.1 \pm 1.2)\times 10^{-2} $  $[17 \%]$ &  $ (1.3 \pm 0.5)\times 10^{-2} $  $[7 \%]$ \\
\hline
\end{tabular}
\tablecomments{Stacked intensities in units of $\rm nW\, m^{-2}\, sr^{-1}$. Uncorrected intensities are labeled \lq\lq Stack\rq\rq, completeness-corrected values are labeled \lq\lq CC\rq\rq, and the best-fit SED values are labeled \lq\lq SED\rq\rq.  Blank spaces represent where the algorithm fails to adequately converge. Errors estimated with a Monte Carlo Simulation described in \S~\ref{sec:errors}.}
\label{tab:cib_masses_sf}
\end{table*}

%% file: t5.tex
\begin{table*}
\footnotesize
 \centering
\caption{Stacked Intensities of Quiescent Galaxies in Bins of Stellar Mass}
 \begin{tabular}{cccccccccc}
 \hline
 \hline
 Band &  Type & $ {\rm log}(M/{\rm M}_{\odot}) = 9.0$--$10.0 $  & $ {\rm log}(M/{\rm M}_{\odot}) = 10.0$--$11.0 $  & $ {\rm log}(M/{\rm M}_{\odot}) = 11.0$--$12.0 $  \\ 
 ($\micro m$) &  $ $ & ($\rm nW\, m^{-2}\, sr^{-1}$) & ($\rm nW\, m^{-2}\, sr^{-1}$) & ($\rm nW\, m^{-2}\, sr^{-1}$) \\ 
 \hline
\multirow{3}{*}{24} &  Stack &  $ (2.1 \pm 1.1)\times 10^{-3} $  $[0 \%]$ &  $ (4.2 \pm 0.3)\times 10^{-2} $  $[2 \%]$ &  $ (2.2 \pm 0.1)\times 10^{-2} $  $[1 \%]$ \\
& CC &  $ (2.1 \pm 1.6)\times 10^{-3} $  $[0 \%]$ &  $ (5.1 \pm 0.4)\times 10^{-2} $  $[2 \%]$ &  $ (2.2 \pm 0.1)\times 10^{-2} $  $[0 \%]$ \\
& SED &  $ (8.4 \pm 0.0)\times 10^{-3} $  $[0 \%]$ &  $ (5.5 \pm 1.9)\times 10^{-2} $  $[2 \%]$ &  $ (2.0 \pm 0.8)\times 10^{-2} $  $[0 \%]$ \\
\multirow{3}{*}{70} &  Stack &  $ (1.5 \pm 2.5)\times 10^{-2} $  $[0 \%]$ &  $ (1.4 \pm 6.5)\times 10^{-2} $  $[0 \%]$ &  $ (1.9 \pm 2.0)\times 10^{-2} $  $[0 \%]$ \\
& CC &  $ (1.8 \pm 4.4)\times 10^{-2} $  $[0 \%]$ &  $ (3.7 \pm 11.0)\times 10^{-2} $  $[1 \%]$ &  $ (1.9 \pm 2.0)\times 10^{-2} $  $[0 \%]$ \\
& SED &  $ (2.2 \pm 0.0)\times 10^{-2} $  $[0 \%]$ &  $ (1.4 \pm 0.5)\times 10^{-1} $  $[3 \%]$ &  $ (5.0 \pm 2.0)\times 10^{-2} $  $[1 \%]$ \\
\multirow{3}{*}{100} &  Stack &  $ (3.7 \pm 1.3)\times 10^{-2} $  $[0 \%]$ &  $ (1.7 \pm 0.3)\times 10^{-1} $  $[1 \%]$ &  $ (6.5 \pm 1.0)\times 10^{-2} $  $[0 \%]$ \\
& CC &  $ (3.8 \pm 2.3)\times 10^{-2} $  $[0 \%]$ &  $ (2.2 \pm 0.5)\times 10^{-1} $  $[1 \%]$ &  $ (6.6 \pm 1.0)\times 10^{-2} $  $[0 \%]$ \\
& SED &  $ (3.1 \pm 0.0)\times 10^{-2} $  $[0 \%]$ &  $ (2.0 \pm 0.7)\times 10^{-1} $  $[1 \%]$ &  $ (7.1 \pm 2.9)\times 10^{-2} $  $[0 \%]$ \\
\multirow{3}{*}{160} &  Stack &  $ (4.5 \pm 6.5)\times 10^{-2} $  $[0 \%]$ &  $ (1.7 \pm 1.8)\times 10^{-1} $  $[1 \%]$ &  $ (1.0 \pm 0.5)\times 10^{-1} $  $[1 \%]$ \\
& CC &  $ (2.1 \pm 1.2)\times 10^{-1} $  $[1 \%]$ &  $ (2.0 \pm 2.9)\times 10^{-1} $  $[1 \%]$ &  $ (1.0 \pm 0.5)\times 10^{-1} $  $[0 \%]$ \\
& SED &  $ (4.4 \pm 0.0)\times 10^{-2} $  $[0 \%]$ &  $ (2.7 \pm 0.9)\times 10^{-1} $  $[2 \%]$ &  $ (1.1 \pm 0.4)\times 10^{-1} $  $[0 \%]$ \\
\multirow{3}{*}{250} &  Stack &  $ (1.8 \pm 1.1)\times 10^{-2} $  $[0 \%]$ &  $ (1.3 \pm 0.3)\times 10^{-1} $  $[1 \%]$ &  $ (1.1 \pm 0.1)\times 10^{-1} $  $[1 \%]$ \\
& CC &  $ (2.1 \pm 1.9)\times 10^{-2} $  $[0 \%]$ &  $ (3.2 \pm 0.5)\times 10^{-1} $  $[2 \%]$ &  $ (1.1 \pm 0.1)\times 10^{-1} $  $[0 \%]$ \\
& SED &  $ (5.4 \pm 0.0)\times 10^{-2} $  $[0 \%]$ &  $ (2.8 \pm 1.0)\times 10^{-1} $  $[2 \%]$ &  $ (1.3 \pm 0.6)\times 10^{-1} $  $[1 \%]$ \\
\multirow{3}{*}{350} &  Stack &  $ (2.2 \pm 1.0)\times 10^{-2} $  $[0 \%]$ &  $ (1.4 \pm 0.3)\times 10^{-1} $  $[3 \%]$ &  $ (9.0 \pm 0.9)\times 10^{-2} $  $[2 \%]$ \\
& CC &  $ (5.8 \pm 1.7)\times 10^{-2} $  $[0 \%]$ &  $ (3.2 \pm 0.5)\times 10^{-1} $  $[3 \%]$ &  $ (9.0 \pm 0.9)\times 10^{-2} $  $[1 \%]$ \\
& SED &  $ (5.6 \pm 0.0)\times 10^{-2} $  $[0 \%]$ &  $ (2.4 \pm 0.9)\times 10^{-1} $  $[2 \%]$ &  $ (9.8 \pm 3.9)\times 10^{-2} $  $[1 \%]$ \\
\multirow{3}{*}{500} &  Stack &  $ (1.1 \pm 0.6)\times 10^{-2} $  $[0 \%]$ &  $ (7.2 \pm 1.7)\times 10^{-2} $  $[3 \%]$ &  $ (5.0 \pm 0.6)\times 10^{-2} $  $[2 \%]$ \\
& CC &  $ (4.1 \pm 1.2)\times 10^{-2} $  $[0 \%]$ &  $ (1.6 \pm 0.3)\times 10^{-1} $  $[3 \%]$ &  $ (5.0 \pm 0.6)\times 10^{-2} $  $[1 \%]$ \\
& SED &  $ (4.9 \pm 0.0)\times 10^{-2} $  $[1 \%]$ &  $ (1.3 \pm 0.5)\times 10^{-1} $  $[2 \%]$ &  $ (4.6 \pm 1.7)\times 10^{-2} $  $[0 \%]$ \\
\multirow{3}{*}{1100} &  Stack &  $ (4.6 \pm 5.1)\times 10^{-4} $  $[0 \%]$ &  $ (2.2 \pm 1.3)\times 10^{-3} $  $[3 \%]$ &  $ (2.4 \pm 0.5)\times 10^{-3} $  $[3 \%]$ \\
& CC &  $ (4.7 \pm 8.9)\times 10^{-4} $  $[0 \%]$ &  $ (6.6 \pm 2.9)\times 10^{-3} $  $[3 \%]$ &  $ (2.5 \pm 0.5)\times 10^{-3} $  $[1 \%]$ \\
& SED &  $ (7.5 \pm 0.0)\times 10^{-3} $  $[4 \%]$ &  $ (9.1 \pm 3.3)\times 10^{-3} $  $[5 \%]$ &  $ (3.1 \pm 1.1)\times 10^{-3} $  $[1 \%]$ \\
\hline
\end{tabular}
\tablecomments{Stacked intensities in units of $\rm nW\, m^{-2}\, sr^{-1}$. Uncorrected intensities are labeled \lq\lq Stack\rq\rq, completeness-corrected values are labeled \lq\lq CC\rq\rq, and the best-fit SED values are labeled \lq\lq SED\rq\rq.  Blank spaces represent where the algorithm fails to adequately converge. Errors estimated with a Monte Carlo Simulation described in \S~\ref{sec:errors}.}
\label{tab:cib_masses_passive}
\end{table*}